\documentclass[sigconf]{acmart}

\AtBeginDocument{%
  \providecommand\BibTeX{{%
    \normalfont B\kern-0.5em{\scshape i\kern-0.25em b}\kern-0.8em\TeX}}}

\copyrightyear{2024}
\acmYear{2024}
\setcopyright{acmlicensed}\acmConference[CHI '24]{Proceedings of the CHI Conference on Human Factors in Computing Systems}{May 11--16, 2024}{Honolulu, HI, USA}
\acmBooktitle{Proceedings of the CHI Conference on Human Factors in Computing Systems (CHI '24), May 11--16, 2024, Honolulu, HI, USA}
\acmDOI{10.1145/3613904.3642902}
\acmISBN{979-8-4007-0330-0/24/05}

\usepackage{booktabs}
\usepackage{graphicx}
\usepackage[utf8]{inputenc}
\usepackage{tabularx}
\usepackage{array}
\usepackage[shortlabels]{enumitem}
\usepackage{array}
\newcolumntype{L}[1]{>{\raggedright\arraybackslash}>{\raggedright}p{#1}}
\begin{document}

\title{The Metacognitive Demands and Opportunities of Generative AI}

\author{Lev Tankelevitch}
\authornote{Both authors contributed equally to this research.}
\affiliation{%
  \institution{Microsoft Research}
  \city{Cambridge}
  \country{United Kingdom}
}
\email{lev.tankelevitch@microsoft.com}

\author{Viktor Kewenig}
\authornotemark[1]
\authornote{The work was done when the co-author was employed at Microsoft.}

\affiliation{%
  \institution{University College London}
  \city{London}
  \country{United Kingdom}
}
\email{ucjuvnk@ucl.ac.uk}

\author{Auste Simkute}
\authornotemark[2]
\affiliation{%
  \institution{University of Edinburgh}
  \city{Edinburgh}
  \country{United Kingdom}}
\email{a.simkute@sms.ed.ac.uk}

\author{Ava Elizabeth Scott}
\authornotemark[2]
\affiliation{%
  \institution{University College London}
  \city{London}
  \country{United Kingdom}
}
\email{ava.scott.20@ucl.ac.uk}

\author{Advait Sarkar}
\affiliation{%
  \institution{Microsoft Research}
  \city{Cambridge}
  \country{United Kingdom}
  }
\email{advait@microsoft.com}

\author{Abigail Sellen}
\affiliation{%
  \institution{Microsoft Research}
  \city{Cambridge}
  \country{United Kingdom}}
\email{asellen@microsoft.com}

\author{Sean Rintel}
\affiliation{%
  \institution{Microsoft Research}
  \city{Cambridge}
  \country{United Kingdom}}
\email{serintel@microsoft.com}

\renewcommand{\shortauthors}{Tankelevitch and Kewenig, et al.}

\begin{abstract}
Generative AI (GenAI) systems offer unprecedented opportunities for transforming professional and personal work, yet present challenges around prompting, evaluating and relying on outputs, and optimizing workflows. We argue that metacognition—the psychological ability to monitor and control one’s thoughts and behavior—offers a valuable lens to understand and design for these usability challenges. Drawing on research in psychology and cognitive science, and recent GenAI user studies, we illustrate how GenAI systems impose metacognitive demands on users, requiring a high degree of metacognitive monitoring and control. We propose these demands could be addressed by integrating metacognitive support strategies into GenAI systems, and by designing GenAI systems to reduce their metacognitive demand by targeting explainability and customizability. Metacognition offers a coherent framework for understanding the usability challenges posed by GenAI, and provides novel research and design directions to advance human-AI interaction.
\end{abstract}

\ccsdesc[500]{Human-centered computing~HCI theory, concepts and models}
\ccsdesc[300]{Human-centered computing~User centered design}
\ccsdesc[500]{Computing methodologies~Artificial intelligence}
\ccsdesc[300]{Human-centered computing~Interaction design theory, concepts and paradigms}

\keywords{Generative AI, Metacognition, Human-AI interaction, User Experience Design, System Usability}

\maketitle

\section{Introduction}\label{sec:intro}
Generative artificial intelligence (GenAI) systems—using models, like Large Language Models (LLMs), that can generate artefacts by using extensive parameters and training data to model and sample from a feature space \cite{bommasani_opportunities_2022}—have the potential to transform personal and professional work. Their potential stems from a unique combination of \textit{model flexibility} (in their input/output space), \textit{generality} (in their applicability to a wide range of tasks), and \textit{originality} (in their ability to generate novel content) \cite{schellaert_your_2023}. However, these same properties also pose a challenge for designing GenAI systems to be human-centered \cite{chen_next_2023}. User studies reveal a range of usability challenges around prompting \cite{zamfirescu-pereira_why_2023}, evaluating and relying on outputs \cite{sarkar_what_2022}, and deciding on an automation strategy: whether and how to integrate GenAI into workflows \cite{barke_grounded_2023,sarkar_exploring_2023}. 

Recent work has sought to characterize the unique properties of GenAI and their potential effects on users \cite{schellaert_your_2023, sarkar_what_2022}, and to offer technical or design roadmaps for designing human-centered GenAI \cite{chen_next_2023,weisz_toward_2023}. However, there is not yet a coherent understanding of the usability challenges of GenAI, much less one grounded in a theory of human cognition. Indeed, recent work has called for foundational research to understand how people interact with GenAI and AI more broadly \cite{lai_towards_2023,liao_ai_2023}. Here, we argue that \textit{metacognition}—the psychological ability to monitor and control one’s own thought processes \cite{norman_metacognition_2019,fiedler_metacognition_2019,ackerman_meta-reasoning_2017,tarricone_taxonomy_2011}—offers a valuable and unexplored perspective to understand and design for the usability challenges of GenAI. Firstly, we suggest that current GenAI systems impose multiple metacognitive demands on users; understanding these demands can help interpret and probe the identified and potentially novel usability challenges. Secondly, we suggest that the perspective of metacognitive demands offers new research and design opportunities for human-AI interaction. 

The metacognitive demands of working with GenAI systems parallel those of a manager delegating tasks to a team. A manager needs to clearly understand and formulate their goals, break down those goals into communicable tasks, confidently assess the quality of the team's output, and adjust plans accordingly along the way. Moreover, they need to decide whether, when, and how to even delegate tasks in the first place. Among others, these responsabilities involve the metacognitive \textit{monitoring} and \textit{control} of one's thought processes and behavior \cite{nelson_metamemory_1990,fiedler_metacognition_2019,ackerman_meta-reasoning_2017,tarricone_taxonomy_2011}. 

Analogously, current GenAI systems often require verbalized prompting, demanding self-awareness of task goals, and decomposition of tasks into sub-tasks. System outputs then need to be evaluated, requiring well-adjusted confidence in one’s evaluation and prompting abilities, and metacognitive flexibility to iterate on the prompting strategy as necessary. Alongside the local interactions with GenAI systems, the generality of GenAI poses another, higher-level metacognitive demand: the challenge of knowing whether and how to incorporate GenAI into workflows---i.e., one’s `automation strategy' (see also \cite{sarkar_exploring_2023}). This demands self-awareness of GenAI’s applicability to, and impact on, one’s workflow; well-adjusted confidence in manual versus GenAI-supported task completion; and metacognitive flexibility to adapt one’s workflows as needed. We posit that these metacognitive demands are induced by GenAI's model flexibility, generality, and originality.\footnote{Although the perspective of metacognition is equally relevant for understanding the usability of search engines and similar technologies, we focus our scope to GenAI systems as they are relevantly distinct from that of search engines \cite{schellaert_your_2023}. Firstly, they are more flexible in their responsiveness to user prompts, in the range of implicit and explicit parameters available to users in their prompts, and in the multi-modality of their input/output space. Secondly, unlike search engines, they function as general-purpose tools, able to perform content generation, discrimination, and editing, among other functions (rather than merely retrieve existing content). Finally, unlike search engines, current GenAI systems are non-deterministic in their responses. As we aim to demonstrate here, all of these features place unique demands on users' metacognition and inform the design space of solutions to address these demands, a design space which necessarily extends beyond that of current search engines. Relatedly, we also note that Russell \cite{design_lab_what_2017} proposed a connected idea of `meta-literacy’ for search engine usability (see also \cite{mackey_reframing_2017}); however, this work does not delve into the psychological and cognitive science of metacognition that is central to the current work.} In §\ref{sec:demands}, we draw on metacognition research and recent user studies of GenAI to illustrate these metacognitive demands and offer new research directions to probe them further.

These demands can be addressed in at least two complementary ways. Firstly, given that metacognitive abilities can be taught \cite{muijs_metacognition_2020, donker_effectiveness_2014,de_boer_long-term_2018}, we can \textit{improve users’ metacognition} via metacognitive support strategies that can be integrated into GenAI systems. Evidence-based metacognitive support strategies include those that help users in their planning, self-evaluation, and self-management \cite{schunk_self-regulation_2000}. Recent HCI work has begun to pursue this direction \cite{suh_sensecape_2023, wu_ai_2022}, albeit without explicitly grounding it in metacognition; we suggest that a metacognitive lens offers new research and design directions for augmenting GenAI system usability. 

Secondly, we can \textit{reduce the metacognitive demand} of GenAI systems by designing task-appropriate approaches to GenAI explainability and customizability. We suggest that explainability can help offload metacognitive processing from the user to the system, and that existing explainability approaches can be augmented by considering metacognition. Likewise, we suggest that a metacognitive perspective can provide insights on approaching the end-user customizability of GenAI systems. In §\ref{sec:addressing}, we draw on intervention studies to improve metacognition and studies of GenAI prototypes and human-AI interaction to explore research and design directions that can address the metacognitive demands of GenAI. Critically, we also highlight how GenAI’s model flexibility, generality, and originality can serve as a design solution to these demands. Finally, we discuss the relationship between cognitive load and addressing metacognitive demands, offering ways to manage their balance. In summary, our work makes three distinct contributions:
\begin{enumerate}
    \item We conceptualize and ground the usability challenges of GenAI in an understanding of human metacognition, drawing on research from psychological and cognitive science and recent GenAI user studies.
    \item We draw from research on metacognitive interventions, GenAI prototypes, and human-AI interaction to propose two directions for addressing the metacognitive demands of GenAI: improving users’ metacognition, and reducing the metacognitive demands of GenAI. 
    \item We use the metacognition lens to identify the need—and concrete directions—for further research into the metacognitive demands of GenAI, and design opportunities that leverage the unique properties of GenAI to augment system usability.
\end{enumerate}
In the next sections, we define metacognition, summarizing key research findings (§\ref{sec:whatismc}); illustrate the metacognitive demands of GenAI, focusing on prompting, evaluating and relying on outputs, and deciding on one’s automation strategy (§\ref{sec:demands}); and propose ways to address these metacognitive demands (§\ref{sec:addressing}). 

\section{What is metacognition?}\label{sec:whatismc}

Metacognition as a concept was first popularized by developmental psychologist John H. Flavell in the late 1970s \cite{flavell_metacognition_1979}, as he tried to understand how children come to be aware of their own cognitive processes. Subsequently, Nelson and Narens \cite{nelson_metamemory_1990} showed that while adults are able to reflect on their thoughts, they often fail to be aware of the premises underlying their decision-making, and do not analyze, understand, and control their thought processes objectively. Their %
`metacognitive model' first distinguished between object-level and meta-level cognition. \textit{Object}-level processes reflect the basic cognitive work of perceiving, remembering, classifying, deciding, and so on. \textit{Meta}-level processes monitor those object-level processes to assess their functioning (e.g., assessing how well one grasped the gist of a text) and allocate resources appropriately (e.g., deciding to re-read the text). Since then, a growing line of research has linked improved metacognition to a range of benefits across different domains. Studies have shown that improved metacognition helps individuals with management of time, focus, and effort \cite{zimmerman_theories_2001}, problem-solving \cite{georghiades_general_2004}, academic performance \cite{livingston_metacognition_2003,tanner_promoting_2012,zohar_review_2013, donker_effectiveness_2014, de_boer_long-term_2018,muijs_metacognition_2020}, emotional well-being \cite{wells_metacognitive_2009}, and overall decision-making \cite{yeung_metacognition_2012}. 

As we argue in §\ref{sec:demands}, alongside the promises of GenAI to transform work, it also poses usability challenges that can be fruitfully understood via metacognition. Nevertheless, the field of human-computer interaction (HCI) has so far considered metacognition mainly in the context of computer science education \cite{prather_what_2020,loksa_metacognition_2022}. The relative absence of metacognition research from many areas of HCI is surprising, considering that the early work on graphical user interfaces was, as Alan Kay concluded, \textit{``solidly intertwined with learning''} \cite{kay_user_1990}. One possible reason for this absence is the confusing plethora of existing and overlapping frameworks and theories on metacognition. From education \cite{livingston_metacognition_2003} to management \cite{keith_self-regulation_2005}, healthcare \cite{church_how_2023}, and even sports \cite{macintyre_metacognition_2014}, many research disciplines have carved out their own approach to metacognition, producing multiple inconsistent terminologies and frameworks (for reviews see \cite{tarricone_taxonomy_2011, norman_metacognition_2019}). 

\begin{figure*}[h]
\centering
  \includegraphics[width=1\linewidth]{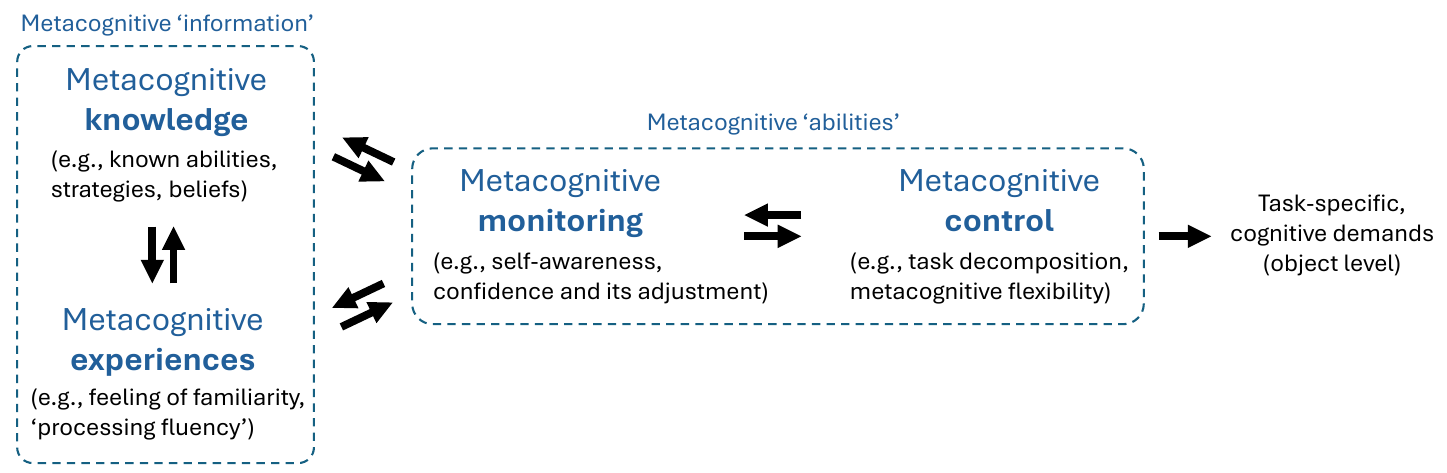}
  \caption[]{A simplified descriptive framework for %
  metacognition. Metacognitive \textit{knowledge} is the explicit understanding of one's abilities, strategies, and beliefs. Metacognitive \textit{experiences} include things that people can directly experience, such as a feeling of familiarity or other implicit cues that provide information about cognitive processes. Metacognitive knowledge and experiences are interrelated in that experiences can become encoded as knowledge, and knowledge can be retrieved during experiences (§\ref{subsec:MCinter}). Both of these can influence (and be influenced by) metacognitive \textit{monitoring}, which includes self-awareness, and confidence and its adjustment. Metacognitive monitoring, in turn, influences (and is influenced by) metacognitive \textit{control} processes, such as metacognitive flexibility and task decomposition. Metacognitive control acts upon the (object-level) cognitive processes involved in a task. Arrows indicate directions of influence (§\ref{subsec:MCinter}). %
  }
  \Description{The image is a flow diagram describing the components of metacognition, divided into `Metacognitive information' and `Metacognitive abilities'. On the left side under `Metacognitive information', there are two interconnected components. The top component is labeled `Metacognitive knowledge' and includes examples like known abilities, strategies, and beliefs. Below it is a component labeled `Metacognitive experiences', with examples such as feeling of familiarity and `processing fluency'. In the center, under `Metacognitive abilities', interconnected to both `Metacognitive knowledge' and `Metacognitive experiences', is a component labeled `Metacognitive monitoring'. This includes examples like self-awareness, confidence, and its adjustment. On the right side, interconnected to `Metacognitive monitoring', is a box labeled `Metacognitive control'. Examples provided include task decomposition and metacognitive flexibility. Finally, an arrow leads from `Metacognitive control' to the right, pointing to a term outside of the box labeled `Task-specific, cognitive demands (object level)'. The diagram illustrates a process flow where metacognitive information feeds into metacognitive monitoring (and vice versa), which in turn informs metacognitive control (and vice versa), ultimately affecting how one handles task-specific cognitive demands.}
  \label{fig:Figure 1}
\end{figure*}

To structure our analysis of the metacognitive demands of GenAI systems, in §\ref{subsec:MCknowexp}-\ref{subsec:MCdomgenspec} we present a simplified descriptive framework of metacognition, also summarized in \autoref{fig:Figure 1}. In line with most common prior frameworks, we distinguish between metacognitive \textit{knowledge} and \textit{experiences}, two different sources of information for understanding one's own cognition \cite{tarricone_taxonomy_2011, norman_metacognition_2019,efklides_metacognition_2008}, and between the metacognitive abilities of \textit{monitoring} and \textit{control}, through which one can assess and guide their own cognition \cite{fiedler_metacognition_2019, ackerman_meta-reasoning_2017, nelson_when_1991}.

\subsection{Metacognitive knowledge and experiences}\label{subsec:MCknowexp}
\textit{Metacognitive knowledge}, being explicit, includes people's conscious understanding of aspects like their strategies (e.g., memory strategies \cite{nelson_metamemory_1990}), reasoning abilities, decision-making, and beliefs \cite{stanovich_individual_2000}.  

\textit{Metacognitive experiences} include anything that people can directly experience, and can be implicit, occurring without our direct intention or awareness \cite{efklides_metacognition_2008}. This includes subjective feelings, like a feeling of familiarity, or the feeling that one has misunderstood a passage while reading, as well as other implicit cues that provide information about cognitive processing (e.g., `processing fluency' cues, such as the speed at which a memory is retrieved) \cite{norman_metacognition_2019,ackerman_meta-reasoning_2017}. 

Metacognitive knowledge and experiences are interrelated \cite{flavell_metacognition_1979}. Metacognitive experiences can contribute to metacognitive knowledge—e.g., when feelings of difficulty during problem-solving become encoded as knowledge that one is poor at problem-solving. Metacognitive knowledge can also be retrieved during metacognitive experiences, for example, when one remembers that they are poor at problem-solving when experiencing a feeling of difficulty. 

\subsection{Metacognitive abilities: monitoring and control}\label{subsec:MCabilities}
\textit{Monitoring} abilities involve the assessment of one's own thinking, whereas \textit{control} abilities are those that directly guide one's own thinking. Our focus is on %
the monitoring and control abilities that %
are most relevant to concrete task-oriented metacognitive demands posed by GenAI (%
see \cite{tarricone_taxonomy_2011} for a more in-depth taxonomy%
).

Relevant monitoring abilities for working with GenAI include self-awareness and adjustment of confidence. \textit{Self-awareness} is the capacity to recognize one’s own thoughts, emotions, and actions, as well as how these factors influence cognition \cite{gravill_metacognition_2002, zimmerman_theories_2001}. This includes having a clear awareness of one’s specific goals and intentions—for example, \textit{``What am I trying to convey with this email?''}. This ability is important for prompting GenAI and determining one's automation strategy (§\ref{subsec:prompting} and §\ref{subsec:automation}). 

\textit{Confidence} is one’s self-assessment of one’s cognitive abilities and their application to tasks \cite{yeung_metacognition_2012}—e.g., \textit{``How confident am I that I can write this email with the appropriate tone and level of detail?''} A `well-adjusted' confidence distinguishes objectively correct and incorrect performance, and accurately matches one’s abilities.\footnote{While beyond the scope of this work, metacognition research distinguishes between two formal and independent aspects of confidence: \textit{resolution} (also known as sensitivity), the ability for confidence judgments to distinguish correct and incorrect performance, and \textit{calibration} (also known as bias), the extent to which confidence tends to be overall higher or lower than objective performance \cite{fleming_how_2014,fleming_metacognition_2023}. We indicate this distinction in relevant points, but direct interested readers to the cited work for more information.} Confidence and its adjustment are central to decision-making and reasoning, especially in many aspects of human-AI interaction \cite{ackerman_meta-reasoning_2017,steyvers_three_2023} (§\ref{subsec:prompting}, §\ref{subsec:evaluating}, and §\ref{subsec:automation}).

Relevant control abilities for working with GenAI include metacognitive flexibility and task decomposition. \textit{Metacognitive flexibility} is the ability to adaptively shift cognitive strategies when encountering new information, when realizing that a current strategy isn’t effective, or when the demands of the task change \cite{canas__cognitive_2005}---e.g., \textit{``I recognize that my formal tone in my emails does not match the more conversational style of my new co-workers. I should therefore adjust my approach while still maintaining professionalism''}. It is a hallmark of creative problem-solving \cite{preiss_metacognition_2022} and has been deemed essential for organizing and integrating a rapidly changing body of information \cite{meltzer_teaching_2014}. Metacognitive flexibility is especially important when prompting and evaluating the output of GenAI (§\ref{subsec:prompting} and §\ref{subsec:evaluating}) and determining one's automation strategy (§\ref{subsec:automation}).

\textit{Task decomposition} involves breaking down a task into concrete, actionable sub-tasks or steps. For instance, before writing an email, one might set clear objectives for what specific points to communicate---e.g., \textit{``I want to clearly explain the status of the project and ask for feedback.''} Then, one might decide on a structure for the email, laying out the most important aspects first. These abilities are especially important for prompting GenAI (Section \ref{subsec:prompting}). As the example suggests, task decomposition is not solely metacognitive because it often involves object-level cognitive processes. 

Monitoring and control are interrelated \cite{koriat_metacognition_2007, koriat_intricate_2006}. Monitoring (i.e., assessing our performance) affects control (e.g., by influencing a change in strategies). Control (e.g., changing strategies) can also provide feedback which affects monitoring (e.g., by altering the assessment of our performance). 

\subsection{Interrelationship between knowledge and experiences, and monitoring and control}\label{subsec:MCinter}

Metacognitive knowledge and experiences interact with monitoring and, in turn, with control \cite{tarricone_taxonomy_2011}. For example, adequate metacognitive knowledge of the strengths and weaknesses of one’s strategies can affect the adjustment of confidence in a solution to a problem (i.e., metacognitive monitoring). Conversely, improving monitoring, such as by practicing self-awareness, can increase one’s awareness of metacognitive experiences and knowledge. Likewise, metacognitive experiences can influence, and be influenced by, our metacognitive monitoring and control. For example, after experiencing a sense of misunderstanding, we might unconsciously adjust our sense of confidence (monitoring), and be prompted to re-read a passage (control). Similarly, the impact of metacognitive experiences might vary based on monitoring abilities. For example, a person with better monitoring might be more attuned to these experiences (thereby also making them less implicit) \cite{zimmerman_theories_2001}. Thus, there is a tight interrelationship between metacognitive knowledge and experiences, and monitoring and control.\footnote{Some metacognition theories view metacognitive knowledge and experiences not as separate from monitoring, but rather as \textit{instances} of monitoring that can be either knowledge- or experience- based \cite{ackerman_meta-reasoning_2017, nelson_metamemory_1990,koriat_metacognition_2007}. However, we distinguish between the two sets of concepts to emphasize the difference between the \textit{ability} to monitor and control cognition, and \textit{information} about cognition arising from metacognitive knowledge and experiences—e.g., the difference between the ability to be self-aware about one's memory (monitoring) and the information conveyed by a feeling of familiarity (an experience).} 

\subsection{Domain generality and specificity of metacognition}\label{subsec:MCdomgenspec}
Whether metacognitive abilities, knowledge, and experiences are \textit{domain-general} (universally applicable across different areas of knowledge, skills, or problem-solving) or \textit{domain-specific} (pertain to a particular area of expertise to which they are finely tuned, such as math) is a matter of active debate, although there is evidence for both views \cite{gilbert_strategic_2015, rouault_human_2018, de_beukelaer_changing_2023, azevedo_reflections_2020,mazancieux_towards_2023}. What kind of metacognition a situation demands is likely context-dependent \cite{efklides_metacognition_2008, de_beukelaer_changing_2023}, a particularly relevant consideration for GenAI given its generality across domains \cite{schellaert_your_2023}. 

\subsection{Heuristics and priming metacognition}\label{subsec:MCpriming}
People often implicitly and unintentionally rely on heuristics to guide their %
metacognitive monitoring and control \cite{ackerman_heuristic_2019}. For example, people guide their metacognitive control by implicitly relying on the ease of information processing (`processing fluency').
Information that is easily or fluently processed (e.g., in terms of reading) triggers less further cognitive processing relative to information that is more difficult to process \cite{ackerman_heuristic_2019}. %
Because processing fluency is subjectively experienced rather than consciously known, and because these heuristics are often activated 
implicitly, their activation represents a metacognitive experience \cite{ackerman_heuristic_2019, unkelbach_general_2013}.\footnote{In contrast, the conscious \textit{knowledge} of these heuristics exemplifies metacognitive knowledge.} %
Manipulating the activation of these heuristics (e.g., via priming) can be used to improve metacognitive control: increasing subjective processing difficulty (e.g., by using a degraded font) stimulates more metacognitive control and thereby improves participants’ performance on reasoning tasks that benefit from a more analytic processing style \cite{alter_overcoming_2007}. In §\ref{sec:demands}, we discuss how %
engaging these heuristics %
when interacting with GenAI %
may influence users. 

\subsection{Improving metacognition}\label{subsec:MCimprove}
Metacognitive abilities can be taught and improved \cite{donker_effectiveness_2014, de_boer_long-term_2018}. Metacognitive interventions, such as training through feedback on metacognitive performance \cite{carpenter_domain-general_2019}, can, for example, increase judgment accuracy—the ability to distinguish between one’s own correct and incorrect metacognitive processing. Other interventions include providing feedback to adjust a person’s mental model for a specific task \cite{weis_know_2022}, and using guided reflection to improve the ability to discern the reliability of  outputs \cite{tanner_promoting_2012}. In §\ref{subsec:improvingmc}, we discuss how some of these interventions can be used in practice to meet the metacognitive demands posed by GenAI. 

\subsection{Measuring metacognition}\label{subsec:MCmeas}
Metacognition researchers have devised a range of methods to measure different metacognitive abilities, both prospectively and retrospectively%
. \autoref{tab:measures} summarizes %
key methods from metacognition research relevant to %
exploring interactions with GenAI.

\begin{table*}
\small
  \caption[]{Overview of some prospective and retrospective methods for exploring relevant metacognitive abilities. Using these methods to measure the metacognitive demands posed by GenAI and applying them for improving GenAI usability are promising opportunities for future research (see \autoref{tab:rqs-demands} and \autoref{tab:rqs-opportunities}).}
  \label{tab:measures}
\renewcommand{\arraystretch}{1.4}  \begin{tabular}{>{\raggedright}p{58pt}>{\raggedright}p{48pt}>{\raggedright}p{108pt}>{\raggedright}p{240pt}}
    \toprule
    \textbf{Ability} & \textbf{Type} & \textbf{Measure} & \textbf{Description} \cr
    \midrule
    Self-awareness & Prospective & Think-aloud \cite{ericsson_protocol_1993} & Users verbalize their thought process during a task. \cr
    & Prospective & Self-report \cite{grant_self-reflection_2002} & Users report their perceived strengths and weaknesses. \cr
    & Prospective & Prediction log \cite{efklides_trends_2010} & Users predict performance and feelings for an upcoming task. \cr
    & Retrospective & Reflective essay \cite{hatton_reflection_1995} & Users describe their thought processes after task completion. \cr
    & Retrospective & Interview \cite{kvale_interviews_1994} & Interviews focus on users' self-perception during or after a task. \cr
    & Retrospective & Assessment rubric \cite{andrade_using_2000} & Users assess their performance using a predefined rubric. \cr
    \hline
    Confidence & Prospective & Judgment of learning (JOL) \cite{nelson_when_1991} & Users predict their performance before a test. \cr
    & Prospective & Self-rating \cite{bandura_self-efficacy_1997-1} & Users rate their confidence in specific skills before a trial or task. Correlation between confidence and objective accuracy can estimate confidence \textit{calibration} \cite{nelson_comparison_1984, winne_chapter_2000}; \textit{Meta-d'} is a derived metric capturing users' prospective confidence \textit{resolution} independent of their calibration \cite{maniscalco_signal_2014,fleming_how_2014} (see also \cite{fleming_metacognition_2023, rahnev_measuring_2023}). \cr
    & Prospective & Likelihood estimate \cite{yeung_metacognition_2012} & Users estimate the likelihood of success in a future event. \cr
    & Retrospective & Self-Rating \cite{bong_academic_2003} & Users rate their confidence in their performance after a trial or task. \cr
    & Retrospective & Reflective journals \cite{moon_j_learning_2000} & Users reflect and comment on how confident they felt during the task. \cr
    \hline
    Task decomposition & Prospective & Expectancy questionnaire \cite{eccles_motivational_2002} & Users set specific goals and plans before a task. \cr
    & Prospective & Self-regulated learning (SRL) microanalysis \cite{cleary_self-regulation_2004} & Users respond to prompts assessing strategy use and motivational beliefs. \cr
    & Prospective & Goal-setting worksheet \cite{zimmerman_self-regulation_2009} & Users fill out a survey about a task's value and expected success. \cr
    & Retrospective & Performance reviews \cite{segedy_designing_2015} & Users evaluate their self-regulation strategies used in a task. \cr
    & Retrospective & Behavioural observations \cite{perry_talking_2008} & Recorded task sessions are coded for indicators of self-regulated learning. \cr
    & Retrospective & Reflective interview \cite{kvale_interviews_1994} & Interviews explore users’ strategic planning, monitoring, and evaluation. \cr
    \hline
    Metacognitive flexibility & Prospective & Cognitive flexibility scale \cite{martin_new_1995} & Users describe how they solve problems in different contexts. \cr
    & Prospective & Task switching \cite{monsell_task_2003} & Users are tested on their ability to switch between task sets. \cr
    & Prospective & Category fluency \cite{troyer_clustering_1997} & Users list examples within categories in a given time. \cr
    & Retrospective & Post-task debrief \cite{spiro_cognitive_1991} & Interviews about users' different strategies used and adaptability during the task. \cr
    & Retrospective & Solution review \cite{jonassen_instructional_1997} & Users review and discuss the solutions they generated for a task. \cr
    & Retrospective & Error analysis \cite{carter_anterior_2007} & Mistakes made during task performance are analyzed to understand metacognitive flexibility. \cr
  \bottomrule
  \end{tabular}\renewcommand{\arraystretch}{1}
\end{table*}

\section{The metacognitive demands of generative AI}\label{sec:demands}

As \citet{sarkar_what_2022} notes, programming with GenAI may have \textit{``far-reaching impact on [programmers’] attitudes and practices of authoring, information foraging, debugging, refactoring, testing, documentation, code maintenance, learning, and more''}. Other domains, such as design \cite{gmeiner_exploring_2023}, writing \cite{noy_experimental_2023}, and data science \cite{gu_how_2023} are likely to be experiencing similar changes with GenAI. We suggest that a core dimension underlying these changes is a greater demand on users’ metacognition that is imposed when users have to (a) prompt GenAI systems, (b) evaluate and decide to rely on GenAI output, and (c) decide on one’s workflow automation strategy: whether they should automate certain tasks with GenAI and how to automate them most effectively (previously described as \textit{``critical integration''} \cite{sarkar_exploring_2023}; see also \cite{risko_thinking_2023}). 

It is important here to distinguish between \textit{metacognitive demand}—the need for extensive metacognitive monitoring and control for a task—and \textit{cognitive load}, the total amount of mental effort required for a task \cite{sweller_cognitive_1998}. Metacognitive demand contributes to cognitive load, but so do other aspects related to cognitive processing at the object (non-meta) level (i.e., metacognitive demand is sufficient but not necessary for increasing cognitive load). For example, as we illustrate below, prompting in current GenAI systems imposes a high metacognitive demand due to the need for\break self-awareness of goals, increasing cognitive load, while the interaction method of typing (rather than speaking) further increases cognitive load, albeit without much associated metacognitive demand. The relationship between metacognitive demand and cognitive load becomes relevant when considering interventions to support users' metacognition (see §\ref{subsec:cogload}).\footnote{For an in-depth theoretical discussion of metacognition (or the overlapping concept of self-regulated learning) and cognitive load, see \cite{seufert_interplay_2018,de_bruin_synthesizing_2020,schwonke_metacognitive_2015,valcke_cognitive_2002}.}    

\begin{figure*}[h]
\centering
  \includegraphics[width=\linewidth]{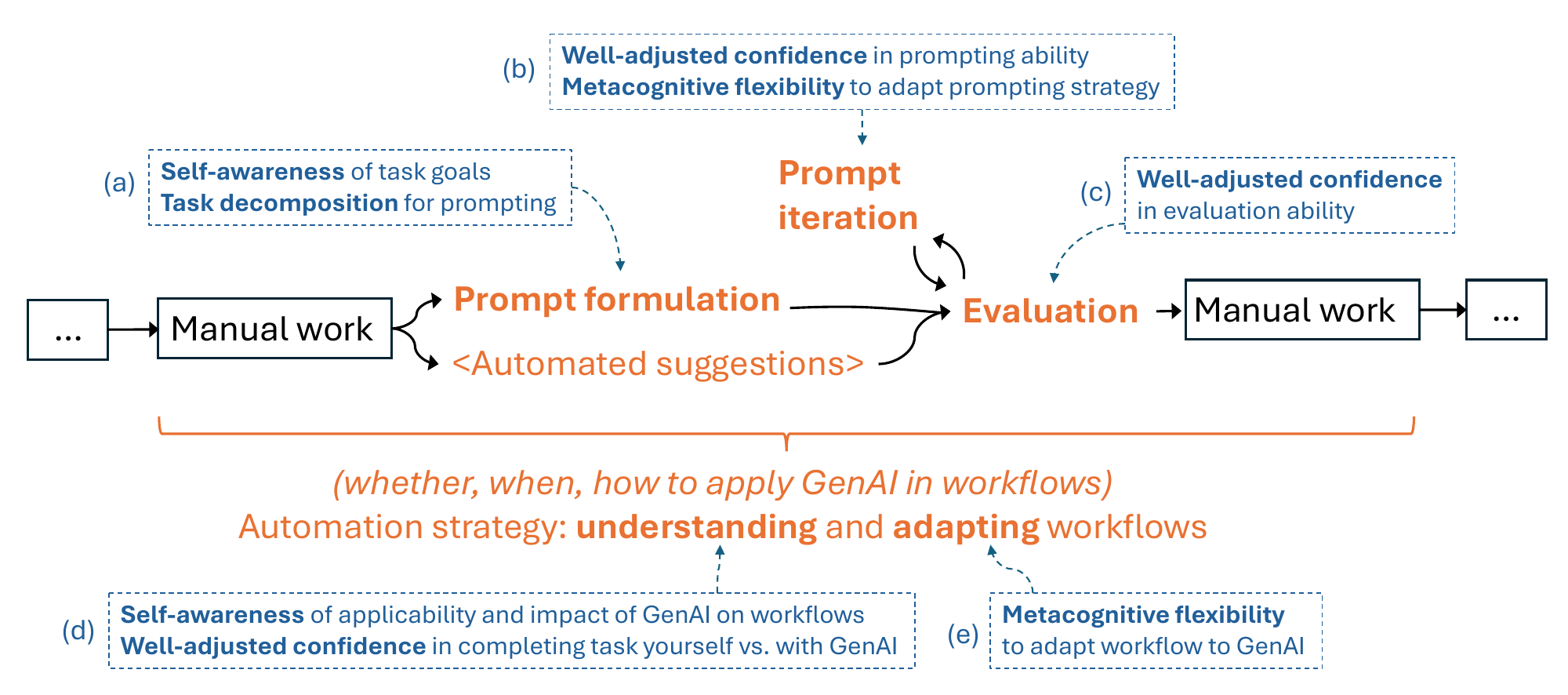}
  \caption[]{Metacognitive demands posed by generative AI at each point in a simplified user workflow. Often embedded within a workflow with manual tasks, users may first need to formulate a prompt, requiring metacognitive abilities including self-awareness of task goals and task decomposition (a). Systems that provide automated suggestions such as GitHub Copilot alleviate some of the demands associated with prompting. Depending on the output, iterating on the prompt may be necessary, which requires well-adjusted confidence in one's prompting ability and metacognitive flexiblity to adapt prompting strategies as necessary (b). Likewise, evaluating the output requires well-adjusted confidence in one's ability to judge its validity (c). Beyond the local interaction with a GenAI system, there is an overarching demand connected to understanding whether, when, and how to apply GenAI to one's workflows---one's `automation strategy'. This requires self-awareness of how GenAI applies to and affects one's workflows, and well-adjusted confidence in the ability to complete tasks manually and with GenAI (d). Finally, it also requires metacognitive flexbility to adapt one's workflows as necessary (e).}
  \Description{The image is a flowchart that illustrates a typical simplified workflow which integrates manual work with generative AI. Starting from the left, there is a sequence of processes indicated by ellipses leading to `Manual work'. This manual work flows into `Prompt formulation', which is influenced by two factors: (a) `Self-awareness of task goals' and `Task decomposition for prompting', and (b) `Well-adjusted confidence in prompting ability' and 'Metacognitive flexibility to adapt prompting strategy'. From `Prompt formulation', the process moves into `Prompt iteration', which is a loop with `Evaluation'. The `Evaluation' is influenced by (c) `Well-adjusted confidence in evaluation ability'. After evaluation, the process either returns to `Prompt iteration' or moves forward to `Manual work' followed by the continuation of the process indicated by ellipses. Beneath this flow, there is a bracket summarizing the overall strategy as `Automation strategy: understanding and adapting workflows', which includes whether, when, how to apply GenAI in workflows. Below the bracket, there are additional influencing factors: (d) `Self-awareness of applicability and impact of GenAI on workflows' and `Well-adjusted confidence in completing task yourself vs. with GenAI', and (e) `Metacognitive flexibility to adapt workflow to GenAI'. This flowchart depicts the metacognitive considerations involved in effectively incorporating AI into work processes, with an emphasis on self-awareness, confidence, and flexibility.}
  \label{fig:Figure 2}
\end{figure*}

This section covers each aspect of working with GenAI systems, describing how GenAI imposes high metacognitive monitoring and control demands on users (summarized in \autoref{fig:Figure 2}). Not all GenAI systems impose the same type and extent of metacognitive demands due to differences in interface design and interaction modes; where relevant, we point out the implications of this. Throughout, we make concrete suggestions on future research to better understand these demands (summarized in \autoref{tab:rqs-demands}).  

\subsection{Prompting generative AI systems}\label{subsec:prompting}
End-user studies suggest that prompting is challenging, with non-expert users making various errors and adopting ineffective strategies—a reflection of the demand on users' metacognitive monitoring and control \cite{chen_machine_2023,dang_how_2022,dang_choice_2023,xu_-ide_2022, jiang_discovering_2022,liang_understanding_2023,sun_investigating_2022}. During \textit{prompt formulation}, the open-endedness of many current prompting interfaces requires users to have self-awareness of their specific task goals, and be able to decompose their tasks into smaller sub-tasks so as to verbalize these as effective prompts (\autoref{fig:Figure 2}a). Next, iterative output evaluation and adjustment (\textit{prompt iteration}) depends on users’ confidence in their prompting ability, and metacognitive flexibility to adapt their prompting strategy (\autoref{fig:Figure 2}b). We posit that these demands are exacerbated by GenAI's non-determinism and model flexibility (not to be confused with metacognitive flexibility) in terms of (a) the wide range of explicit and implicit parameters that users can adjust, and (b) systems’ ability to work with prompts at a wide range of abstraction \cite{sarkar_what_2022,schellaert_your_2023}.\footnote{A popular workaround to the challenge of prompting is `prompt libraries' with detailed, task-specific prompts (see \cite{svendsen_outline_2023} for an overview). While helpful, ready-made prompts will rarely suit one’s context precisely—the devil remains in the details. Moreover, ready-made prompts still require metacognitive ability to apply appropriately.}

\subsubsection{Prompt formulation: self-awareness and task decomposition}\label{subsubsec:prompting1}
In manual task completion, many implicit goals and intentions embedded within tasks can remain so without ever being verbalized. For example, when writing an email to a senior colleague, one might implicitly know to adopt a certain tone. Many GenAI systems require specification that the email is to a senior colleague and needs an appropriate tone. Moreover, it often requires that a task be broken up into sub-tasks (\textit{``combine my content'', ``condense into two paragraphs'', ``update the tone''}). This demand for self-awareness and task decomposition is exacerbated by a particular type of model flexibility in GenAI: today’s systems afford many parameters for end-users to adjust; these can be formal parameters like the model temperature, or a range of unspecified parameters that can be adjusted through text prompting (e.g., the tone, level of detail, or structure of a piece of text). This model flexibility and control afforded to users requires knowing what one wants to achieve and convey that explicitly and effectively to the system. Recent user studies of GenAI systems illustrate these demands. 

In \cite{zamfirescu-pereira_why_2023}, non-expert participants used an LLM-based tool to improve a chatbot through prompting. One of the challenges they experienced was a struggle getting started.\footnote{Support getting started is a key user request for GenAI explainability; see §\ref{subsubsec:explain1}.} \citet{zamfirescu-pereira_why_2023} interpret this as a design-stage barrier in end-user programming, reflecting some version of the implicit question,\textit{ ``I don’t even know what I want the computer to do''}. The self-awareness and explicitness demanded by prompting is also observed in LLM-supported writing. \citet{dang_choice_2023} define and compare \textit{diegetic} prompting (instructions conveyed implicitly when users input text for the system to modify) and \textit{non-diegetic} prompting (explicit instructions to the system). The latter is experienced as far more challenging by users as it \textit{``forces writers to shift from thinking about their narrative or argument to thinking about instructions to the system''} \cite{dang_choice_2023}. Similar difficulty with non-diegetic prompting was observed among novice programmers in AI-assisted coding \cite{jayagopal_exploring_2022}, and manufacturing designers co-creating with GenAI, who struggled to \textit{``think through the design problem in advance''} \cite{gmeiner_exploring_2023}. These difficulties were exacerbated by the many parameters available to users who grappled with understanding and using them effectively \cite{gmeiner_exploring_2023,jayagopal_exploring_2022, wu_ai_2022}. 

The difference between diegetic and non-diegetic prompting points to the broader question around GenAI system interfaces and interaction modes and what they imply for the user experience and for productivity. For example, whereas diegetic prompting is easier, it affords less control than the alternative, with preferences differing across users \cite{dang_choice_2023}. Moreover, user experience and productivity may not always go hand in hand. For example, systems with non-diegetic prompting (e.g., ChatGPT) may be more challenging and time-consuming, yet the explicitness they require may plausibly act as a forcing function that ultimately trains metacognitive self-awareness and task decomposition, leading to higher quality output, assuming users persevere \cite{sarkar_should_2023}.\footnote{The potential discrepancy between user experience and productivity is reminiscent of that found in education, where the cognitive effort of effective learning is experienced negatively by students, leading to a divergence between \textit{perceptions} of the learning experience and objective learning outcomes%
.} Nevertheless, in §\ref{subsec:improvingmc} we suggest that there are more effective and user-friendly ways of supporting metacognition.   

Apropos of training, a key difference between expert and non-expert programmers---and by extension, expert and non-expert prompt writers---is an explicit approach to considering task requirements \cite{zamfirescu-pereira_why_2023}. Expert programmers have advanced metacognitive monitoring and control, in that they are able to identify their specific goals and decompose them into concrete tasks \cite{etelapelto_metacognition_1993}. One developer in \cite{liang_understanding_2023} described their strategy with LLM-supported coding as, \textit{``be incredibly specific with the instructions and write them as precisely as I would for a stupid collaborator''}. Likewise, users in \cite{barke_grounded_2023} who decomposed the programming task into ``\textit{microtasks}''—``\textit{well-understood and well-defined jobs}''—were able to work effectively with Copilot (see also \cite{ross_programmers_2023, vaithilingam_expectation_2022}). Beyond coding, manufacturing designers who successfully learned to co-create with GenAI abstracted and explained the problem to themselves \cite{gmeiner_exploring_2023}.\footnote{As the above quotes suggest, task decomposition can often mean crafting a prompt as a set of discrete instructions that a system can interpret all at once, but it can also require step-wise prompting, which can work sequentially to produce the desired output, or, in some cases, may require manual reassembly of multiple outputs.}

We note that, although self-awareness and task decomposition are often required to some extent when interacting with GenAI systems, their pertinence increases as users concretize their usage intentions (e.g., achieving work or personal goals). For example, users interacting with GenAI systems for non-specific entertainment or exploration may worry less about their prompting strategy. At the same time, systems that support users' metacognition can help surface or clarify intentions originally hidden from users' self-awareness, thereby influencing their initial goals or lack thereof (see §\ref{subsubsec:selfeval} for details). For example, users `playing' with a system may be enabled to identify more concrete or diverse forms of play for them to explore. Thus, our framework of metacognitive demands is applicable to many use-cases. More generally, we do not assume that users' intentions and goals (or lack thereof) remain static during human-AI interaction, and, as per §\ref{subsubsec:selfeval}, suggest that systems can and should help users clarify their intentions and goals.

Future research should systematically examine how self-awareness and task decomposition ability moderate users' ability to control systems across interaction modes (e.g., diegetic vs. non-diegetic prompting), task contexts (e.g., creating a novel output vs. editing an existing artifact), and domains (e.g., writing vs. programming). 

\subsubsection{Prompt iteration: confidence adjustment and metacognitive flexibility}\label{subsubsec:prompting2}
After the initial prompt, the next common step is iteration: evaluating the output and adjusting the prompt accordingly (here we focus on prompting; see §\ref{subsec:evaluating} on evaluating the output). Alongside maintaining awareness of their task goals, users need to (\autoref{fig:Figure 2}b):
\begin{enumerate}[(a)]
    \item evaluate the output with respect to their prompt,
    \item adjust their confidence in their prompting ability, to disentangle this from systems’ capabilities (\textit{``Is my prompt specific and clear enough; are system parameters set appropriately; is system performance generally poor on this task; or is this an `unlucky' probabilistic output?''}),
    \item flexibly adjust their prompting strategy as needed (\textit{``Should I adjust my prompt, adjust an earlier prompt, decompose my tasks into further sub-tasks, re-try with the same prompt…etc.?''}).
\end{enumerate}
The range of possible explanations for a poor output makes confidence adjustment challenging, and the range of possible strategy adaptations demands high metacognitive flexibility from the user. This is exacerbated by the non-determinism of GenAI, particularly when tweaking one aspect of the prompt might unintentionally change a different aspect of the output \cite{chen_machine_2023}. This requires constantly maintaining awareness of one’s task goals in the face of ever-changing output, or risk getting derailed from the task by unexpected output, as some users were in \cite{wu_ai_2022}.\footnote{The usability challenges of prompting make it a key target for explainability, as per §\ref{subsubsec:explain1}.}  

It is further exacerbated by a distinct type of model flexibility in GenAI systems (not to be confused with metacognitive flexibility): \textit{``[generative AI systems] can generate plausible and correct results for statements at an extremely wide range of abstraction''}, which presents what \citet{sarkar_what_2022} term a `fuzzy abstraction matching' problem—it becomes difficult for users to discern a system’s capabilities and to match one’s intent and prompting accordingly (see also \cite{jiang_discovering_2022,ferdowsifard_small-step_2020} for similar conclusions).

Participants in \cite{zamfirescu-pereira_why_2023} struggled with this (see also \cite{dang_how_2022,jiang_discovering_2022,xu_-ide_2022}). They were unable to choose the right prompting instructions, incorrectly expecting human capabilities; in some cases, underestimating the system’s capabilities; and insisted on socially appropriate—rather than effective—ways of prompting. Zamfirescu-Pereira et al. interpret these challenges as stemming from \textit{``over-generalization from limited experience''}, and \textit{``a social lens that filtered participants’ prompts…through expectations originating in human-human interactions''}. From a metacognitive perspective, over-generalization reflects poorly adjusted confidence; rather than maintaining an appropriately low confidence (about their own prompting), and gathering more evidence, participants drew confident conclusions based on limited evidence. Participants’ insistence on a social lens may reflect a lack of self-awareness of their prompting approach, poorly adjusted confidence, and/or inflexibility in their strategies. To be clear, these challenges partly stem from a lack of feedback in the system about prompting effectiveness, leaving users to grapple with the fuzzy abstraction matching problem without support (see also §\ref{subsubsec:explain1} on explainability). However, that participants in \cite{zamfirescu-pereira_why_2023} \textit{``avoided effective prompt designs even after their interviewer encouraged their use and demonstrated their effectiveness''} suggests that this is also partly a metacognitive failure to notice and/or adjust their mental model of the system, signaling low metacognitive flexibility (see also \cite{wu_ai_2022, gmeiner_exploring_2023}).

Future research should systematically investigate how different aspects of GenAI systems—such as their non-determinism and model flexibility—impact users’ ability to adjust their confidence in their prompting ability, flexibly adapt their prompting strategy, and update their mental model of these systems. For example, this could examine how the temperature setting of a model influences users’ confidence and its adjustment, or how different levels of abstraction influence novice users’ prompting strategies.   

\subsection{Evaluating and relying on generative AI outputs}\label{subsec:evaluating}
Evaluating and relying on AI output requires users to maintain a well-adjusted confidence in their own domain expertise and ability to evaluate output (i.e., self-confidence; \autoref{fig:Figure 2}c). The importance of this metacognitive demand is evidenced in recent research on AI-assisted decision-making, which finds that users’ self-confidence is a key determinant of their reliance on AI responses, as discussed in §\ref{subsubsec:evaluating1} below. Confidence in the \textit{system’s} abilities is also important, and likely interacts with self-confidence, although here we focus on the latter as it is a metacognitive concept (i.e., an assessment of one's cognition via metacognitive monitoring).\footnote{Confidence in the system’s abilities is touched upon in §\ref{subsubsec:explain1} on explainability. Note also that self-confidence in output evaluation and confidence in the system's ability are both distinct from users’ self-confidence in their \textit{prompting} ability, discussed above in §\ref{subsec:prompting}. Prompting and output evaluation influence each other as users iterate on their task.}

We posit that GenAI exacerbates the demand for a well-adjusted confidence in output evaluation. In this context, this includes confidence with `good' \textit{calibration}, meaning the overall confidence of a user in their output evaluation accurately matches objective performance; and with `good' \textit{resolution}, meaning the user's confidence can correctly discriminate a correct output.\footnote{The two aspects of confidence can be independent. Having a well-calibrated confidence does not necessarily imply high confidence resolution. One could be well-calibrated on average (e.g., one's overall level of confidence matches one's overall level of accuracy) but still have poor resolution (i.e., one's confidence level does not vary much between correct and incorrect answers)\cite{fleming_how_2014}.} The generative nature of GenAI---its ‘originality’ \cite{schellaert_your_2023}---means that many user workflows will or have shifted from users \textit{generating} content to \textit{evaluating} it \cite{risko_thinking_2023,sarkar_exploring_2023}, as already documented in programming \cite{sarkar_what_2022} and manufacturing design \cite{gmeiner_exploring_2023}. Thus, users must maintain a well-adjusted level of confidence in their own ability to evaluate this output and not blindly accept generated content. Moreover, GenAI poses %
unique challenges to confidence adjustment that we discuss below.

\subsubsection{AI output evaluation and reliance: confidence adjustment}\label{subsubsec:evaluating1}
Recent work has investigated the role of self-confidence in evaluating and relying on AI output, although with discriminative models, rather than GenAI. For example, in AI-assisted decision-making in chess, participants’ reliance on the AI was only significantly predicted by their self-confidence, and not by their confidence in the AI \cite{chong_human_2022}. This dovetails with \citet{he_knowing_2023} who find that poor performers in a logical reasoning task tend to be overconfident---the Dunning-Kruger effect---leading to under-reliance on AI (see \cite{tejeda_ai-assisted_2022} for related findings). \citet{lu_human_2021} show that, in the absence of AI accuracy information, humans rely on their agreement with AI as a heuristic for their reliance on it, albeit only when they have a high self-confidence. Older human-automation interaction studies demonstrated a similar role for user self-confidence in influencing reliance on automation \cite{de_vries_effects_2003,lee_trust_1994,madhavan_similarities_2007}.

Analogous research in GenAI that explicitly measures and manipulates user self-confidence is missing, but user studies suggest a similar key role for well-adjusted confidence in output evaluation and reliance.\footnote{Related to output evaluation, confidence and metacognition have also been studied in the context of phishing detection \cite{canfield_better_2019}.} Programmers were reluctant to deeply review and repair AI-generated code, preferring instead to re-write the entire code themselves \cite{vaithilingam_expectation_2022}. Similarly, manufacturing designers co-creating with GenAI were uncertain about how to interpret outputs and whether users or the system were responsible for addressing errors \cite{gmeiner_exploring_2023}. By contrast, programmers who were highly confident in their own ability actively questioned the AI code assistant when it produced confusing output \cite{weisz_perfection_2021}.

The challenge of output evaluation is present even in interaction modes without user prompting, such as in GitHub Copilot, which includes automated suggestions. In fact, such interaction modes may arguably make output evaluation more challenging due to the need to infer the intent behind systems' suggestions \cite{gu_how_2023}. Novices in a domain or in GenAI may be particularly vulnerable, as one participant commented on long code suggestions: \textit{``if you do not know what you’re doing, it can confuse you more''} \cite{prather_what_2020}. Given the importance of users’ self-confidence for evaluating and relying on AI outputs, future research should measure and manipulate self-confidence during user-GenAI interactions across different interaction modes. 

Output evaluation is relevant for many systems, such as search engines, but several aspects of GenAI pose unique challenges, which we discuss below: the \textit{extensiveness} of GenAI's novel content output, the relative \textit{ease} of novel content generation, GenAI's multiple, non-intuitive \textit{failure modes}, and the challenge of obtaining \textit{objective quality measures} for adjusting confidence in some workflows.

\textit{\textbf{The extensiveness of GenAI's novel content output.}} Whereas prior research has focused on explicit AI advice or decisions, GenAI can produce (often extensive) content, such as entire emails, presentations, or software. Evaluating these outputs for quality therefore becomes far more important and effortful (in terms of cognitive load) compared to `auto-complete' phrase suggestions or intelligent code completion \cite{sarkar_what_2022,schellaert_your_2023}. How this will affect users’ self-confidence and AI reliance  remains unclear, but metacognition research suggests that increased effort requirements may discourage users from appropriate evaluation \cite{ackerman_meta-reasoning_2017}. \citet{ackerman_diminishing_2014} found that people’s internal confidence threshold for solving reasoning problems decreases as the required effort increases; that is, \textit{``when problems took longer to solve, participants appeared to compromise on their confidence criterion, and were willing to provide solutions with less confidence''}. Worryingly, this persists even when participants are given the option to give up and respond ``I don’t know'' \cite{ackerman_diminishing_2014}. Likewise, end-user programmers have been reported to ``eyeball'' the AI outputs of natural language queries, which some suggest may deepen the existing over-confidence that such users have in their programs’ accuracy \cite{sarkar_what_2022,srinivasa_ragavan_gridbook_2022}.

Future research should examine how the effort %
of evaluating GenAI output %
(in terms of %
length or complexity) affects users’ self-confidence in their output evaluation, the accuracy of their evaluation, and their ultimate reliance on GenAI.

\textit{\textbf{The relative ease of novel content generation.}} The relative ease with which GenAI can produce extensive output may also affect output evaluation and reliance via potentially misleading cues that people implicitly rely on to update their confidence and guide their subsequent metacognitive control \cite{ackerman_meta-reasoning_2017}. One relevant type of cue—`processing fluency’, \textit{``the subjective ease of with which a cognitive task is performed''} \cite{ackerman_heuristic_2019,wiley_improving_2016}—can influence people’s confidence in information accuracy. For example, answers to various problems are judged as more correct simply if they are displayed faster to participants after problem descriptions \cite{topolinski_immediate_2010}. It can also affect people’s confidence in their memory: the ease with which a memory is retrieved increases participants’ confidence in their later remembering, even though, objectively, easier retrieval was associated with worse future memory performance \cite{benjamin_mismeasure_1998}.\footnote{The influences of processing fluency cues are examples of metacognitive experiences.} Critically, this effect extends to technology use: faster online information search retrieval increases participants’ confidence in their subsequent memory of that information, despite no apparent causal relation between the two aspects \cite{stone_search_2021}. The mere \textit{use} of technology, such as online information search, can also inflate people’s confidence in their knowledge \cite{dunn_distributed_2021,eliseev_understanding_2023,fisher_harder_2021}. 

Analogously, the ability of GenAI systems to quickly and easily generate extensive content may serve as a cue that misleadingly increases users’ confidence, not only in the output itself, but also in their own ability to evaluate it. More importantly, changes in confidence can affect people’s approach to evaluating GenAI output. By increasing people’s confidence, such cues can affect their metacognitive control, leading people to decrease the effort they invest into further deliberate processing, as measured by thinking time and changes-of-mind \cite{ackerman_meta-reasoning_2017,thompson_role_2013,thompson_analytic_2012}. Confidence similarly influences reliance on external reminders and information-seeking \cite{boldt_confidence_2019,desender_subjective_2018} (see also §\ref{subsubsec:automation2}). 

Future research should systematically investigate how aspects of GenAI output (e.g., the speed at which it’s produced, or its verbal fluency, in the case of text) can serve as cues that influence users’ confidence in the output and their ability to evaluate it, as well as the effort they ultimately invest into evaluation. 

\textit{\textbf{Multiple, non-intuitive failure modes of GenAI}}. Users’ confidence and their ability to adjust it may also be challenged by the fact that GenAI tools can have multiple and often non-intuitive failure modes \cite{chen_machine_2023, schellaert_your_2023}. For instance, they can introduce subtle, non-intuitive errors that a human would not introduce, further complicating evaluation \cite{sarkar_what_2022}. As it stands, this requires developing an expertise and a well-adjusted confidence that is distinct from existing domain expertise, with, for example, \textit{``developers [needing] to learn new craft practices for debugging''} \cite{sarkar_what_2022}. Moreover, as noted, GenAI models are non-deterministic \cite{ouyang_llm_2023}. This is arguably a necessary trade-off within current GenAI systems, as it enables diversity of output \cite{ouyang_llm_2023,sarkar_what_2022}, yet it exacerbates the challenge of confidence adjustment, particularly when working iteratively across prompting and output evaluation (as per §\ref{subsec:prompting}). How confident users should be in their evaluation ability, and how much effort they should invest in evaluation, partly depends on how much non-determinism they can expect in the output. Indeed, manufacturing designers co-creating with GenAI were, \textit{``unable to determine whether…design features were intended or caused by algorithmic glitches''} \cite{gmeiner_exploring_2023}. More broadly, as per §\ref{subsec:prompting}, output failures can be attributed to the user’s prompt or parameter settings, or the system’s non-determinism or training data, without an obvious way to disentangle these, further complicating confidence adjustment, particularly for non-expert users \cite{schellaert_your_2023,weisz_toward_2023}. 

Future work should examine how different reasons for output failures affect users’ ability to appropriately adjust their confidence in their evaluation ability, and how that influences their evaluation of and reliance on GenAI output.

\textit{\textbf{Obtaining appropriate measures for confidence adjustment.}} Adjusting confidence in output evaluation typically requires objective measures of performance for comparison (e.g., the number of errors a user correctly detected in the output), but the quality of generated content and its uses may be more difficult to evaluate objectively (e.g., consider how one would objectively evaluate the quality of an LLM-generated email) \cite{lai_towards_2023, katyal_construct_2023}. The benefits of generated content may also be diffuse and indirect. For example, participants co-writing with an LLM found that seeing the LLM’s suggestions was helpful even when they did not implement them \cite{yuan_wordcraft_2022}. This implies subjectivity in the workflow which, although valid, makes it challenging to adjust one’s confidence. Even with use-cases that involve subjectivity, such as creative tasks, users need to adopt an appropriate reliance strategy, which requires well-adjusted confidence. For example, given the ease of idea generation with GenAI, how can users be confident that the ideas generated by such systems are in fact helpful for their ideation process, rather than merely \textit{feeling} like they are helpful?  

Future work should develop more varied objective measures of output quality, and explore how user-provided subjective measures of quality can support users’ ability to adjust their confidence in output evaluation (e.g., by considering the self-consistency of their reports \cite{katyal_construct_2023}).  

\subsection{Automation strategy and generative AI workflows}\label{subsec:automation}
Beyond the metacognitive demands implied in the local interaction with GenAI, the generality of GenAI—its applicability to a wide range of tasks\cite{schellaert_your_2023}—poses a higher-level question to end-users about their workflow automation strategy: whether they should \textit{``employ Generative AI, how, and how much is the utility of incorporating generated contents compared to conventional approaches''} \cite{chen_next_2023}. \citet{sarkar_exploring_2023} describes this change as a shift from production to `critical integration', where \textit{``the output of AI systems will need to be integrated into a wider workflow involving human action''}, a process requiring critical evaluation of outputs. We argue that this imposes a distinct metacognitive demand on users that must make these decisions, akin to \cite{risko_thinking_2023}, who make a similar argument for digital storage and memory. That is, users must have self-awareness of the applicability and potential impact of using GenAI for their workflow; well-adjusted confidence in the ability to complete a task manually versus with GenAI; and metacognitive flexibility in adapting workflows to GenAI (\autoref{fig:Figure 2}d-e). We first briefly summarize early evidence on how GenAI is impacting user workflows, and then discuss the role of metacognition in users’ workflow automation strategy.   

\subsubsection{Early impact of generative AI on user workflows}\label{subsubsec:automation1}
Research on real-world GenAI workflows, primarily in AI-assisted coding, suggests that tools like GitHub Copilot alter users’ workflows in diverse ways \cite{simkute_ironies_2024}. Although many changes may be positive and related to productivity boosts \cite{dakhel_github_2023}, we focus on the challenges to illustrate the demand for metacognition. In a sample of undergraduate students with programming experience, working with Copilot on realistic programming tasks was perceived to be challenging (although participants still strongly preferred it) \cite{vaithilingam_expectation_2022}. Most relevantly, the generation of a long piece of code, particularly with errors, required participants to switch between coding, reading, and debugging, resulting in a high cognitive load (also reported in \cite{barke_grounded_2023} and, in the domain of AI-assisted programming education, in \cite{prather_its_2023}). Some users in \cite{barke_grounded_2023} felt that Copilot was negatively restructuring their workflow by \textit{``forcing them to jump in to write code before coming up with a high-level architectural design''}. They also reported writing more and differently worded comments %
for Copilot, which they then spent time deleting (unlike comments intended for humans).

Research in other domains points to similar potential challenges. Data scientists highlighted workflow integration as a key lever of control that determined the usefulness of AI assistance \cite{mcnutt_design_2023}. Writing workflows also substantially change with ChatGPT, with user time shifting from rough-drafting to editing \cite{noy_experimental_2023}, although the usability challenges that this brings remain to be explored.

Increased switching costs between automated and manual tasks, and automation-related restructuring of tasks in often unproductive ways have been studied in the human-automation interaction field for decades as the ``ironies of automation'' \cite{bainbridge_ironies_1983,simkute_ironies_2024}. Automated system design has adhered to best practices in human factors engineering to mitigate the impact of these challenges in specific contexts, such as driving. However, current GenAI systems present two key differences: they are applicable to a wide range of tasks (`generality' \cite{schellaert_your_2023}), and as a result, they are also widely available to users with different levels of domain expertise, system training, and workflow standardization, in line with broader automation trends \cite{janssen_history_2019}. Thus, current systems shift the task of managing one’s automation strategy to the user, who may lack expertise or training, leaving them to manage their attention and re-structured workflows as they see fit—a distinct metacognitive demand of GenAI. Broadly, these changes pertain to \textit{understanding} and then \textit{adapting} one's workflows. We discuss each of these below. 

\subsubsection{Understanding one's workflows: self-awareness and confidence adjustment}\label{subsubsec:automation2}
One key question that pertains to users’ automation strategy is \textit{whether} to automate a certain task. Inappropriate reliance on GenAI may result in lost productivity, increased risk of errors, or potential de-skilling \cite{bommasani_opportunities_2022}. Users must therefore have self-awareness of the applicability of GenAI for their workflow, and well-adjusted confidence in their ability to complete the task manually versus with GenAI \cite{sarkar_what_2022} (\autoref{fig:Figure 2}d). Put simply, it requires answering a version of the following questions: \textit{do I know whether an available GenAI system can help my workflow; do I know how to work with it effectively in the context of my workflow; and how confident am I in this knowledge?} \cite{madhavan_similarities_2007}. In end-user programming, this is known as the `attention investment' problem, in which users must conduct a cost-benefit analysis to decide whether the potential attention costs saved from programming a manual task outweigh the attention costs of implementing the program \cite{blackwell_first_2002}.   

Early research suggests that some users, particularly novices in GenAI and/or the task domain, lack sufficient self-awareness and well-adjusted confidence for working effectively with GenAI systems. For example, programming students in \cite{prather_its_2023} repeatedly spent time editing Copilot suggestions before abandoning them and moving on, or tried to coerce Copilot to provide a correct suggestion, two unproductive interaction patterns that suggest potential over-reliance on GenAI. Similarly, some less experienced programmers in \cite{barke_grounded_2023} were particularly excited about Copilot and would over-rely on it before manually attempting any of the tasks themselves. When compared with the relative absence of such interactions among experienced developers (e.g., \cite{barke_grounded_2023,vaithilingam_expectation_2022}), these interaction patterns illustrate a potential lack of metacognitive self-awareness and confidence in managing one’s workflows. However, \citet{kazemitabaar_studying_2023} found no evidence of over-reliance among novice programmers when learning programming using GenAI. 

The above reports are limited to short study contexts, focusing only on programming. Further in-depth research is needed on the impact of GenAI on realistic workflows, including understanding the role of user self-awareness and confidence, particularly for use-cases outside of programming.  

Deciding to rely on GenAI is a form of `cognitive offloading’—the use of tools external to the mind (e.g., calendars), to reduce the cognitive demand of a task (e.g., remembering an event) \cite{risko_cognitive_2016}. With GenAI, although the intent is often (but not always) to produce an external artefact, many cognitive processes that are traditionally involved in such production are at least partly `offloaded' to GenAI, such as ideation, memory retrieval, and reasoning. For example, although users prompt systems with instructions to generate text, it's the systems which often generate ideas, retrieve relevant information, and structure it into arguments. Psychological research has explored how metacognition affects people’s decisions to engage in cognitive offloading, and can therefore inform our understanding of the metacognitive demands pertinent to users’ automation strategy \cite{gilbert_outsourcing_2023,risko_cognitive_2016,scott_metacognition_2023}. Studies find that people’s self-confidence in task performance or knowledge is a strong determinant of their use of external reminders \cite{boldt_confidence_2019,gilbert_strategic_2015,hu_role_2019}, and search for external information \cite{desender_subjective_2018,schulz_dogmatism_2020}. That is, a lower self-confidence in one’s abilities is associated with more cognitive offloading. The above GenAI user studies demonstrate a similar pattern, where less experienced users were more likely to show patterns consistent with over-reliance on tools like Copilot. Critically, studies on cognitive offloading show that even when accounting for people’s objective performance on a task, their \textit{subjective} self-confidence still influences the decision to engage in cognitive offloading \cite{boldt_confidence_2019,gilbert_strategic_2015}. 

Future work should explore how subjective self-confidence relates to users’ automation strategies with GenAI, and how users, particularly novices, can be supported in having increased self-awareness and a well-adjusted confidence to ensure appropriate reliance on GenAI (see also §\ref{sec:addressing}). 

\subsubsection{Adapting one's workflows: metacognitive flexibility}\label{subsubsec:automation3}
Alongside self-awareness and confidence, working with GenAI requires metacognitive flexibility to be able to effectively adapt one's workflow (\autoref{fig:Figure 2}e). For example, users should be able to recognize when and how the use of GenAI interferes with their workflow, resulting in a net productivity loss, and adjust accordingly. As discussed above, the challenges that some Copilot users faced suggest an under-development of this domain-specific metacognitive ability. Conversely, emerging evidence suggests that some experienced users do employ metacognitive flexibility in their workflows with GenAI. For example, some users with prior Copilot experience disable it entirely due to excessive disruption to their workflows \cite{barke_grounded_2023}. In \cite{liang_understanding_2023}, 26 percent of surveyed programmers cite the distracting nature of GenAI suggestions, and 38 percent cite the time-consuming nature of debugging or modifying generated code, as `very important’ reasons for avoiding tools like Copilot. Certain data scientists in \cite{mcnutt_design_2023} similarly expressed skepticism towards AI assistance, particularly for difficult-to-understand generated code. Likewise, many manufacturing designers in \cite{gmeiner_exploring_2023} who struggled with GenAI ultimately avoided it altogether in their process.   

Other users take a more nuanced approach: \textit{```I turned off auto-suggest and that made a huge difference. Now I’ll use it when I know I’m doing something repetitive that it’ll get easily, or if I’m not 100 percent sure what I want to do and I’m curious what it suggests. This way I get the help without having it interrupt my thoughts with its suggestions'''} \cite{sarkar_what_2022}. More broadly, only about 20-30 percent of Copilot suggestions are accepted by users \cite{ziegler_productivity_2022}. 

This evidence also hints at relevant differences between system interfaces and interaction modes that should be considered in the context of metacognitive demands and adapting workflows. Automated suggestions reflect tighter integration between manual work and GenAI, and no metacognitive demands associated with prompting, yet they nevertheless present challenges such as interruptions. In contrast, prompt-based interactions enable more user control over workflows but present metacognitive demands (as per §\ref{subsubsec:prompting1}). User-controlled suggestions may be a middle-ground, but present their own challenges in terms of inferring system intent (as per §\ref{subsubsec:evaluating1}). 
Deciding between these approaches as a user may contribute to the metacognitive demand associated with determining automation strategy (see also \cite{steyvers_three_2023}).     

This is not to suggest that users' approaches to workflow adaptation above are necessarily optimal for productivity, but rather that they reflect self-awareness and confidence in experienced users about how GenAI impacts their workflow, and the exertion of metacognitive flexibility in an effort to change this. As \citet{sarkar_what_2022} conclude, these emerging ad hoc strategies hint at \textit{``a new cognitive burden of constantly evaluating whether the current situation would benefit from LLM assistance''}—a burden that we identify as distinctly metacognitive. Future work should characterize the role of metacognitive flexibility in adapting one's workflows across different GenAI interfaces and interaction modes.

\begin{table*}
\small
  \caption[]{Open research questions for understanding the metacognitive demands of GenAI}
  \label{tab:rqs-demands}
\renewcommand{\arraystretch}{1.4}  \begin{tabular}{>{\raggedright}p{66pt}>{\raggedright}p{270pt}>{\raggedright}p{108pt}}
    \toprule
    \textbf{Area}&\textbf{Research questions}&\textbf{Example measures of metacognition}\cr
    \midrule
    Prompt formulation & How does self-awareness and task decomposition ability moderate users' ability to control systems across interaction modes (e.g., diegetic vs. non-diegetic prompting), task contexts (e.g., creating a novel output vs. editing an existing artifact), and domains (e.g., writing vs. programming)? & Think-aloud, self-report protocols, SRL microanalysis\cr
    \hline
    Prompt iteration & How do different aspects of GenAI systems (e.g., non-determinism, model flexibility) impact users' ability to adjust their confidence in their prompting ability and flexibly adapt their prompting strategy and mental model of GenAI systems? & Prospective self-ratings of confidence, SRL microanalysis\cr
    \hline
    Output evaluation & How does users' self-confidence in a task domain or with GenAI influence their output evaluation and reliance across different interaction modes? & Prospective and retrospective self-ratings of confidence \cr
      & How does the cognitive load associated with evaluating GenAI outputs (e.g., in terms of output length or complexity) affect users' self-confidence, the accuracy of their evaluation, and their ultimate reliance on AI? & Retrospective self-ratings of confidence, meta-\textit{d}' estimates\cr
      & How do aspects of GenAI output (e.g., the speed at which it is produced, its verbal fluency in the case of text) serve as heuristic cues that influence users' confidence in the output and their evaluation ability, as well as the amount of effort they invest into evaluation? & Retrospective self-ratings of confidence, meta-\textit{d}' estimates\cr
      & How do different reasons for output failures affect users' ability to adjust their confidence in their evaluation ability, and how does that influence their strategies for evaluating and relying on GenAI output? & Prospective judgments of learning, retrospective reflective journals\cr
      & What are useful objective measures of quality for long and/or multidimensional outputs, and how can user-provided subjective measures of quality support users' ability to adjust their confidence in output evaluation? & Self-ratings of confidence\cr
      \hline
    Understanding workflows & How does subjective user confidence and self-awareness in a domain and/or in their ability to work with GenAI relate to users' automation strategies with GenAI? & Retrospective self-ratings of confidence, SRL microanalysis\cr
    \hline
    Adapting workflows & What is the role of metacognitive flexibility in adapting one's workflows across different GenAI interfaces and interaction modes? & SRL microanalysis \cr
  \bottomrule
\end{tabular}
\renewcommand{\arraystretch}{1}
\end{table*}

\section{Addressing the metacognitive demands of generative AI}\label{sec:addressing}
The metacognitive demands posed by GenAI can be addressed in two complementary ways: (1) \textit{improving users’ metacognition} via metacognitive support strategies that can be integrated into GenAI systems, and (2) \textit{reducing the metacognitive demand} of GenAI systems by designing task-appropriate approaches to explainability and customizability. The distinction between the two approaches is not clean-cut, yet helps frame the design space.

Multiple lines of evidence suggest that metacognition can be improved, and that individuals who are supported in specific metacognitive monitoring or control abilities can significantly improve their performance in metacognitively demanding tasks \cite{donker_effectiveness_2014,de_boer_long-term_2018}. This applies across different age groups (from children to adults) \cite{cross_developmental_1988,amzil_effect_2013,delclos_effects_1991}, tasks (e.g., lecture comprehension or mathematical reasoning) \cite{ king_improving_1991,butler_strategic_1998,kramarski_enhancing_2003,huff_using_2009}, time-scales (i.e., immediately as well as delayed) \cite{mevarech_immediate_2008}, and learning settings (i.e., solitary as well as social) \cite{palincsar_reciprocal_1984}. As such, interventions to improve the metacognition of users working with GenAI could be one effective way of meeting the demands of these systems. This includes embedding metacognitive support strategies—for example, supporting users’ planning and self-evaluation—directly into GenAI systems. Interventions can be adapted to the metacognitive abilities and GenAI experience of each user to provide the appropriate level of support, making productive use of GenAI’s model flexibility and generality.  

On the other hand, GenAI systems can be designed to reduce their metacognitive demand. One area ripe for this approach is explainability. Designing human-centered explainable AI (HCXAI) has been an important focus in human-AI interaction research \cite{liao_ai_2023,suresh_beyond_2021,ehsan_human-centered_2020}, but the model flexibility, generality, and originality of GenAI systems poses further challenges, as per §\ref{sec:demands}. Yet these same features of GenAI provide an opportunity to support HCXAI, particularly when considering it through the lens of metacognition. Alongside explainability, the customizability of GenAI systems is another lever to reduce metacognitive demand. Current GenAI systems provide many parameters to users, both explicitly (as settings), and implicitly (as prompting strategies). Finding appropriate ways to surface these can reduce metacognitive demand. 

Below, we discuss how to improve users’ metacognition using three types of metacognitive support strategies that can be employed in GenAI systems: planning, self-evaluation, and self-management. After discussing the range of possible strategies for each kind of metacognitive support, we provide a figure showing a hypothetical example of how they might be used in a scenario within an existing GenAI system (analogously to \cite{buschek_how_2022}). We then turn to how systems can be designed to reduce metacognitive demand, focusing on explainability and customizability. We provide examples from research on metacognition interventions and existing prototype studies and suggest opportunities for further research. Lastly, we briefly discuss the importance of managing the cognitive load associated with metacognitive interventions.

\subsection{Improving user metacognition}\label{subsec:improvingmc}

\subsubsection{Planning}\label{subsubsec:planning}
Planning is a task-oriented metacognitive strategy entailing both the definition of clear goals (i.e., self-awareness) and devising a comprehensive approach for achieving them by breaking them down into smaller, manageable steps (i.e., task decomposition) \cite{boekaerts_self-regulation_2005}. Planning-related interventions can support both of these aspects as users work with GenAI systems. 

As discussed below in §\ref{subsubsec:selfeval}, self-evaluation interventions can help users reflect on their task goals and approaches to task decomposition \cite{gu_how_2023}. However, tasks may nevertheless be complex or ambiguous, often requiring gathering, organizing, and synthesizing information in a nonlinear manner distinct from the linear conversational interfaces in most GenAI systems today. For example, people can engage in multi-level planning, where hours, days, and weeks have to be considered simultaneously \cite{ahmetoglu_plan_2021}. More flexible interfaces can support the crafting of a complex prompting strategy by enabling open-ended exploration of task goals and the relationships between them during the planning process. \textit{Sensecape} is such an interface for LLMs that uses multilevel abstraction and visuo-spatial organization to support exploration and sensemaking during LLM interactions \cite{suh_sensecape_2023}. Similar systems have been shown to improve users' planning and other metacognitive processes \cite{crescenzi_supporting_2021}. When applied to the prompting and output evaluation process itself, these approaches can make users aware of where they are in a task. They can also help them encode information in personalized representational schemas \cite{pirolli_sensemaking_2005}, which can help users understand the underlying mechanisms of a GenAI system, its capabilities, and failure points. Such an understanding can in turn mitigate the risks of inappropriate evaluation of, and confidence in, AI-generated output.

Planning-related interventions can also support users directly in task decomposition, improving prompt effectiveness via more explicit and discrete instructions. One promising approach is `prompt chaining', which involves \textit{``decomposing an overarching task into a series of highly targeted sub-tasks, mapping each to a distinct LLM step, and using the output from one step as an input to the next''} \cite{wu_ai_2022,wu_promptchainer_2022}. Alongside improving the LLM’s ability to execute complex tasks, chaining helped participants \textit{``think through the task better''}, and thereby make more targeted edits to improve their prompting \cite{wu_ai_2022}. %
Chaining also increased users’ self-awareness of their goals: the ability to decompose tasks led some participants to create more generalizable outputs better suited to their broader goals \cite{wu_promptchainer_2022}.

 Planning can also help users address the `fuzzy abstraction matching' problem \cite{sarkar_what_2022}, that is, translate their goals and intentions into executable actions—a form of externalization where `tacit' knowledge is made into explicit prompts \cite{nonaka_knowledge-creating_2003}. This can be supported through feedforward design \cite{chen_next_2023} (as distinct from feedback \cite{vermeulen_crossing_2013}): inviting an action and communicating what exactly the user can expect as a result.\footnote{This is also a form of explanation; see §\ref{subsubsec:explain1}.} For example, feedforward can be used to inform users that a vague, high-level prompt is unlikely to achieve their task before they submit it. As \citet{vermeulen_crossing_2013} argue, \textit{``the more complex a system or interaction context gets, the larger the need will be for elaborate feedforward in order to aid users in achieving their goals''}. Prompt chaining is an example of a sequential feedforward approach, surfacing inputs, outputs, and the underlying prompt structure for users to explore \cite{wu_ai_2022}. However, feedforward information can also increase cognitive load \cite{dang_ganslider_2022}, so the right balance, particularly for complex systems, remains to be explored. The \textit{Prompt Middleware} framework uses feedforward at different levels of complexity to guide users towards effective prompts and scaffold domain expertise into the process \cite{macneil_prompt_2023}.

\begin{figure*}[h]
\centering
  \includegraphics[width=\linewidth]{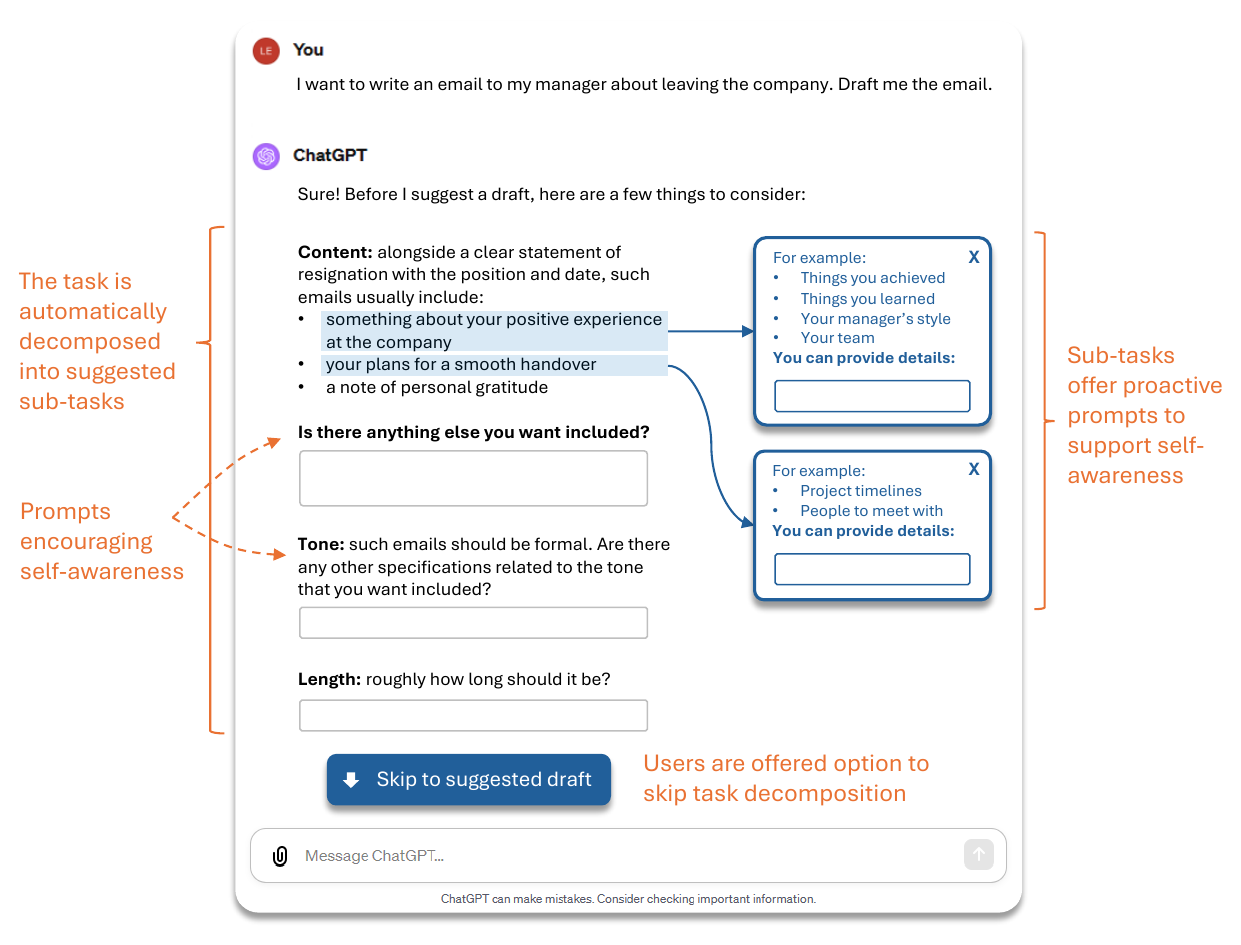}
  \caption[]{Hypothetical example of a planning-focused metacognitive intervention built into ChatGPT. After the user specifies a task, the system automatically comes up with a decomposed, step-by-step guide for completion (left side of the figure). This could be aided by further proactive prompting, giving concrete examples of how sub-tasks could be solved (right side of the figure). An option to skip the decomposition step (bottom of the figure) minimises unnecessary cognitive load if decomposition is not required.}
  \Description{The image depicts a graphical user interface of a chat conversation between a user and ChatGPT, designed to assist with drafting an email to the user's manager about leaving the company. At the top left, there is a box with the user's request to the AI: ``I want to write an email to my manager about leaving the company. Draft me the email.'' The AI responds in the center with a structured approach to drafting the email, breaking down the task into sub-tasks and prompting the user for further information. The response is highlighted with a note stating, ``The task is automatically decomposed into suggested sub-tasks.'' The AI's response includes a bulleted list for the user to consider: Content: clear statement of resignation with the position and date, positive experiences at the company, plans for a smooth handover, and a note of personal gratitude. Tone: formal tone specifications. Length: the desired length of the email. On the right, two pop-up dialog boxes titled ``For example:'' suggest specific details to be included in a text field, such as achievements, things learned, the manager's style, the user's team, project timelines, and people to meet with. These pop-ups are annotated as ``Sub-tasks offer proactive prompts to support self-awareness.'' Below the AI's response, there is a button labeled ``Skip to suggested draft,'' indicating that users have the option to bypass the task decomposition process. The image emphasizes the AI's role in guiding the user through a self-aware approach to task completion, encouraging consideration of various aspects of the task before producing the final output.}
  \label{fig:design1}
\end{figure*}

\autoref{fig:design1} provides a hypothetical example of a planning-focused metacognitive intervention that could be implemented in a conversational interface such as ChatGPT. Rather than requiring an entire chain structure with full control over inputs and outputs, the benefits of chaining could be derived by surfacing a relevant set of key questions to users in a more accessible format.  

 \subsubsection{Self-evaluation}\label{subsubsec:selfeval}
Self-evaluation involves enabling users to reflect on their knowledge, strategies, and performance, and their respective level of confidence in these. Interventions that cue users to reflect on their goals and strategies via question prompts or conversational interfaces have been shown to improve outcomes in the workplace \cite{kocielnik_reflection_2018,meyer_enabling_2021, zargham_understanding_2022}, education \cite{devolder_supporting_2012}, and other domains \cite{bentvelzen_revisiting_2022, reicherts_make_2020}. GenAI systems, with their model flexibility and generality, have the potential to adaptively nudge this kind of self-evaluation at key moments during user workflows, effectively acting as a coach or guide for users \cite{hofman_sports_2023}. 

Self-evaluation can be used to support effective prompting. For example, \citet{gmeiner_exploring_2023} employed human experts to guide designers during their interaction with GenAI systems; users appreciated critical questions that guided self-evaluation and thought this improved prompting. Gmeiner et al. suggest that GenAI systems can proactively offer similar context-aware self-reflection prompts to support users in thinking through problems. Self-evaluation interventions can rely on a range of design elements to support users in clarifying their task goals, including temporal aspects (e.g., considering past tasks or broader project timelines), comparison (e.g., with similar tasks), and discovery (e.g., through re-framing tasks) \cite{bentvelzen_revisiting_2022}.  

Self-evaluation during the output stage can include interventions that surface previous outputs and ask users to `think aloud' about their thought processes, promoting self-awareness in `real time' and supporting users in detecting hallucinations in GenAI outputs \cite{ji_survey_2023}. For example, simple textual prompts promoting self-awareness significantly decreased participants’ susceptibility to incorrect (albeit realistic) information \cite{salovich_misinformed_2021}. Likewise, self-explanation prompts improved students’ accuracy in evaluating their own understanding of information \cite{wiley_improving_2016}. GenAI systems could also probe for and proactively respond to user uncertainty. Even including a `not sure' option for users auditing LLMs, can enable them \textit{``to reflect on the task specification and the appropriateness of the tests considered''} \cite{rastogi_supporting_2023}. 

Self-evaluation is particularly promising for augmenting GenAI explainability, where it can help increase users' receptiveness to explanations that seek to update their mental models of GenAI systems \cite{ackerman_meta-reasoning_2017, steyvers_three_2023} (see §\ref{subsubsec:explain1} for more on explainability). For example, self-reflection probes can ask users to reflect on their mental model and associated confidence \cite{kulesza_tell_2012}. 

GenAI systems also provide opportunity for \textit{interactive} metacognitive support, encouraging self-evaluation and suggesting adjustments to metacognitive strategies based on context \cite{winne_chapter_2000}. For instance, systems could interactively guide users through the steps of a problem rather than simply providing a solution \cite{gmeiner_exploring_2023}. Indeed, the success of pair programming often stems from the extensive verbalization between programmers, and resultant self-evaluation, rather than the division of labor \cite{hannay_effectiveness_2009}. Interactions with GenAI systems could replicate this with a similar level of reflective depth. Enabling users to critically evaluate AI outputs can build self-awareness by forcing users to rationalize their decisions to the GenAI system \cite{sarkar_what_2022}, enhance metacognitive flexibility by providing users with different perspectives on their task \cite{becker_systematic_2023}, and adjust confidence by adapting explanations to the users’ needs and the models' confidence \cite{ma_who_2023}. Moreover, it can also provide critical feedback to GenAI systems themselves.

However, there are also potential pitfalls in interactivity. As per §\ref{subsec:evaluating}, processing fluency can lead to inflated confidence without necessarily improving objective accuracy. The design of intuitive interfaces for GenAI systems might therefore inadvertently give users a misleading sense of competence, increasing the risk of errors (likewise discussed in cognitive psychology \cite{salovich_misinformed_2021}). Designers therefore need to carefully consider how to improve processing fluency without leading to overconfidence, for example, by including periodic checks that challenge users' assumptions or solutions \cite{koriat_illusions_2005}. This speaks to `seamful' design, which %
leads users to pause or reflect on their engagement with technology by emphasizing \textit{``configurability, user appropriation, and revelation of complexity, ambiguity or inconsistency''} \cite{inman_beautiful_2019, weiser_creating_1994, sarkar_should_2023}.

\begin{figure*}[h]
\centering
  \includegraphics[width=\textwidth]{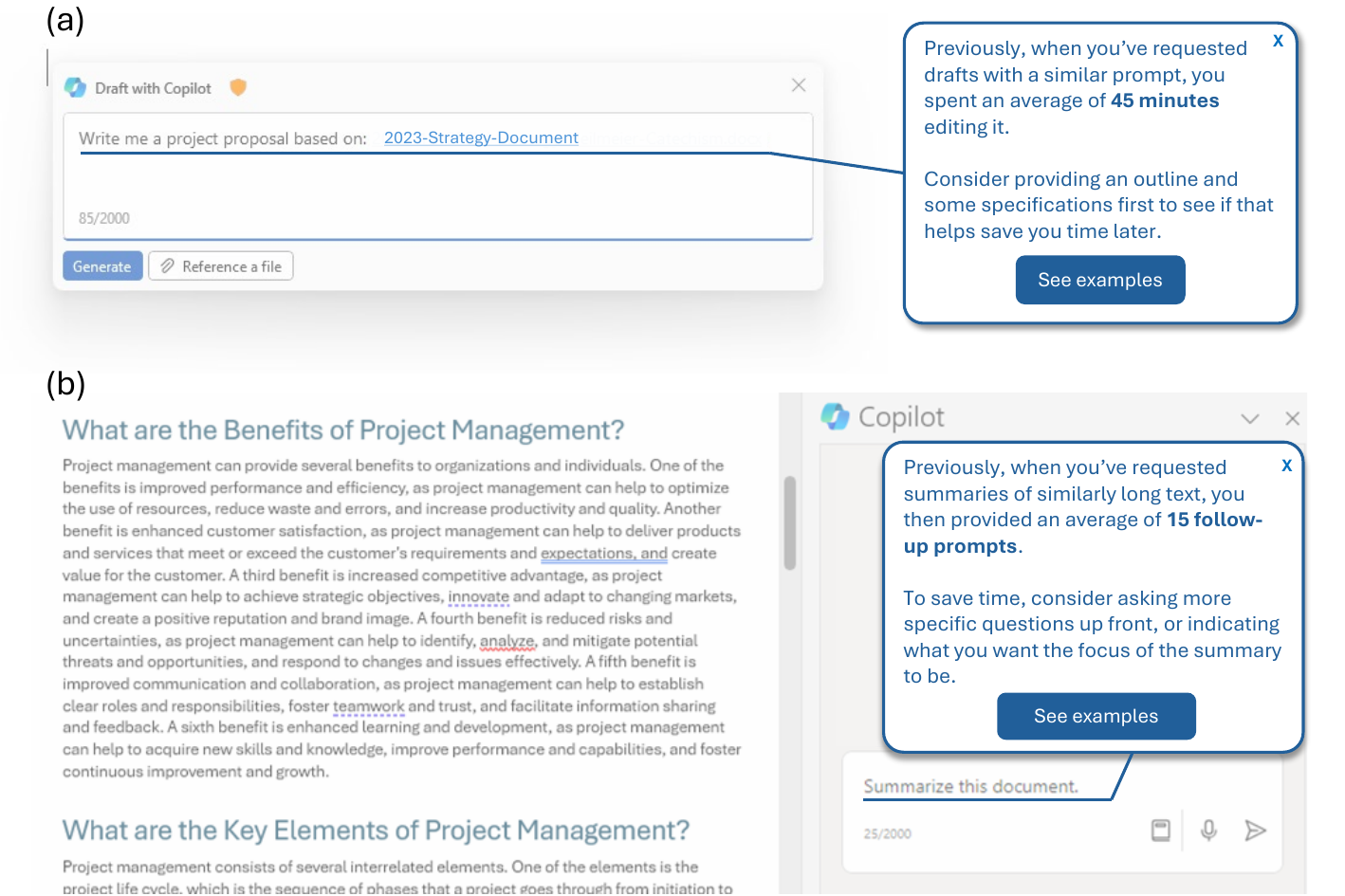}
  \caption[]{Hypothetical example of a metacognitive intervention focused on self-evaluation built into Microsoft Copilot. In (A), the user provides a highly unspecified prompt to the system for writing a proposal. Based on a neutral assessment of similar prompting history, the GenAI system suggests reducing editing time by reflecting on more strategies. In (B) the user provides a highly unspecified prompt for summarizing. Based on a neutral assessment of previous interactions, the GenAI system suggests to limit interactions by suggesting to reflect on the user's more specific goals and intentions for this summary. These appear as suggestions next to the main chat window and can be closed if not wanted.}
  \Description{The image is divided into two components labeled (a), (b), which each display a snippet of a mocked-up user interface in Microsoft Word with a Copilot tooltip. Component (a) shows a screenshot of a Copilot dialog box with text box with a button labeled ``Generate'' and another button to ``Reference a file''. The text box is pre-filled with the prompt ``Write me a project proposal based on: 2023-Strategy-Document''. Adjacent to this box, and pointing to the prompt, is a Copilot tooltip providing historical usage data, suggesting that previously the user spent an average of 45 minutes editing drafts with a similar prompt. It advises considering providing an outline and some specifications first to save time later, with a clickable option to ``See examples''. Component (b) contains a cropped screenshot of faded paragraphs written in Word with a heading ``What are the Benefits of Project Management?'' followed by a paragraph that outlines details. There is a Copilot sidebar to the right with a textboxt pre-filled with a prompt that says ``Summarize this document''. There is a tooltip pointing to this prompt, which suggests that previously the user requested summaries of similarly long text and then provided an average of 15 follow-up prompts. It recommends saving time by asking more specific questions upfront or indicating what the user wants the focus of the summary to be, with a clickable ``See examples'' link.}
  \label{fig:design2}
\end{figure*}

 \autoref{fig:design2} provides a hypothetical example of a metacognitive intervention focused on self-evaluation. To support effective prompting, the user's prior history is leveraged to provide personalized suggestions to improve a generic prompt in this case, and potentially teach the user to include more detail in future prompting.

\subsubsection{Self-management}\label{subsubsec:selfman}
Self-management involves the strategic management of variables like time, setting, and workflow, and is therefore an important focus for metacognitive support strategies for GenAI users. These considerations are not just arbitrary choices but can be informed by a blend of user telemetry trends and explicit user requests. By developing systems that are context-aware, users can be served AI-generated content or prompts at opportune moments during their workflows \cite{gu_how_2023}. For example, coding assistance systems can detect when a user is in a state of flow (`acceleration') or problem-solving (`exploration'), adapting code suggestions accordingly and providing feedback to the user \cite{barke_grounded_2023}. Likewise, during highly sensitive tasks or crucial time periods, the system could trigger heightened user engagement or display more salient reminders to critically evaluate the AI-generated output, promoting self-awareness and adjustment of confidence in output. Another approach is designing the complexity of AI-generated content according to the cognitive load experienced by the user \cite{sweller_cognitive_1988}. This could involve dynamic adaptations, such as recognizing implicit intent \cite{chen_next_2023}, providing summaries when the user is overwhelmed, or escalating the complexity when the user demonstrates high proficiency and engagement. Supporting self-management efficiently also depends on task demands \cite{gu_how_2023}. For example, \cite{abdelshiheed_power_2023} and \cite{shabrina_investigating_2023} suggest that in the context of solving logical puzzles, intelligent tutors should offer backwards-oriented workflows (e.g., using prompts to encourage thinking about the negation of the actual solution), rather than focusing on forward-oriented workflows. In a data science context, \citet{gu_how_2023} propose that GenAI tools can offer \textit{``a `think' mode for specific planning suggestions, a `reflection' mode for connecting decisions and highlighting potential missed steps, and an `exploration' mode for higher-level planning suggestions''}. 

Deciding when to present \textit{any} GenAI support is another design choice affecting self-management. In AI-assisted decision-making, \citet{steyvers_three_2023} distinguish between AI support provision that is on-demand (user-requested) and sequential (occurring after a user makes an independent decision), among others. Apart from facilitating engagement at opportune moments, the sequential paradigm is presumed to encourage independent reflection by the user. \citet{park_slow_2019} similarly argue that ``slower'' interfaces especially enable these benefits, as the waiting time often gets used for reflective thinking about the task at hand (see also \cite{reicherts_make_2020}). Alternatively, workflows can be more dynamic. For example, pathology requires highly specialized, moment-to-moment judgments; in this context, the user capability to control and modify search algorithms on-the-fly can be particularly beneficial \cite{cai_human-centered_2019}.

\begin{figure*}[h]
\centering
  \includegraphics[width=\linewidth]{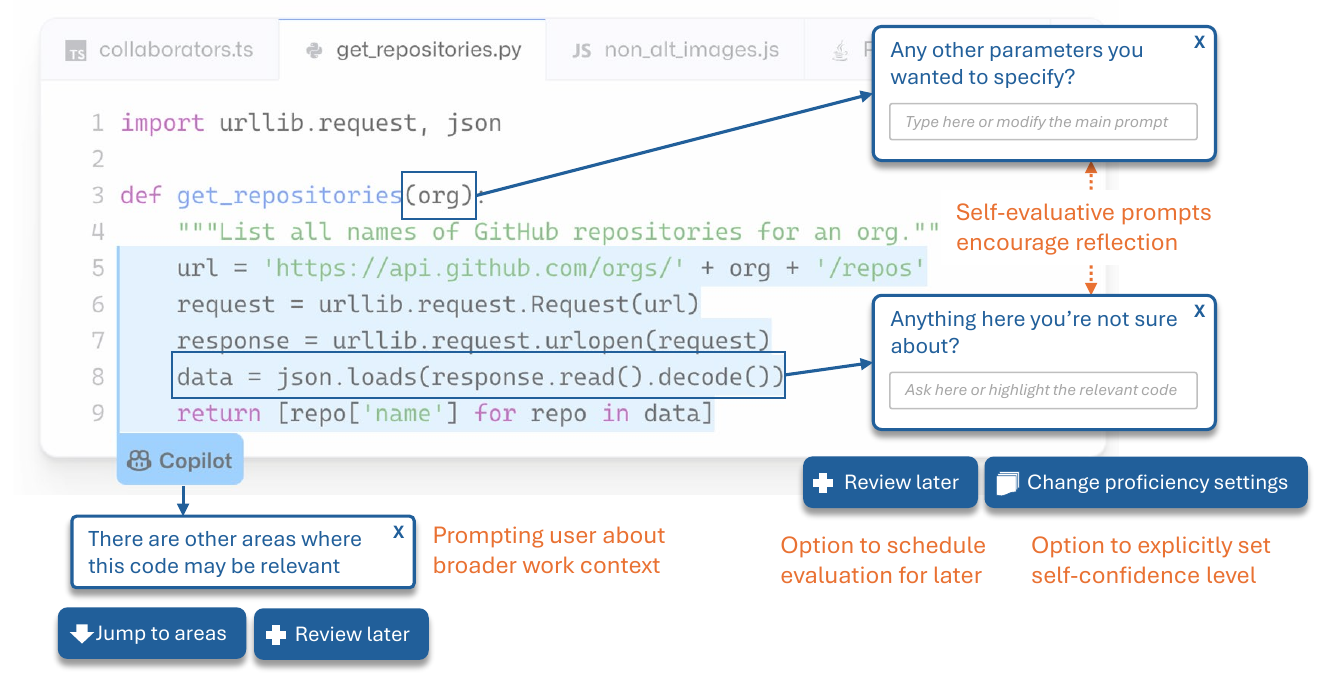}
  \caption[]{Hypothetical example illustrating a metacognitive intervention focused on self-management and self-evaluation for coding in GitHub Copilot. During programming, the system can provide self-evaluation prompts to encourage user reflection on bugs and purpose of the code (right side of the figure). To decrease cognitive load if evaluation is not wanted, the user has the option to schedule evaluation for later, or set their own confidence level to increase or decrease the amount of suggestions (the `proficiency settings' in the bottom right of the figure). The system could also prompt the user to think about the broader work context, for example whether this code-snippet may be relevant somewhere else in the overall code as well. The user has the options to ignore this suggestion, look at the other relevant areas now, or review them later (bottom left of the figure). To minimize cognitive load, the timing and frequency of prompts should be adapted to the users' preferences, expertise, and workflow.}
  \Description{The image is a screenshot of a code editor with an open Python script named get_repositories.py, along with a user interface for an AI coding assistant (GitHub Copilot), providing real-time suggestions and prompts to the coder. The Python script includes an import statement for the urllib.request and json modules and defines a function get_repositories(org) that aims to list all names of GitHub repositories for an organization. The function constructs a URL for a GitHub API call, sends the request, processes the JSON response, and returns a list comprehension that extracts the repository names from the data. Accompanying the code, the AI assistant, labeled as `Copilot', presents several prompts and options. A tooltip points to the function definition and asks, ``Any other parameters you wanted to specify?'' suggesting the coder may need to refine the function. Below the code, a message states, ``There are other areas where this code may be relevant'', with options to ``Jump to areas'' or ``Review later''. Another tooltip provides self-evaluative prompts like ``Anything here you're not sure about?'' to encourage the coder to reflect on their code.
The interface also offers the ability to schedule an evaluation for later, change proficiency settings, and an option to explicitly set the coder’s self-confidence level in their coding abilities. This interface aims to assist the coder by prompting self-awareness and reflection on their code, suggesting additional considerations for their current task, and offering personalized assistance based on their proficiency and confidence levels.}
  \label{fig:design3}
\end{figure*}

\autoref{fig:design3} provides a hypothetical example of a metacognitive intervention focused on self-management and self-evaluation. During coding, a system might encourage the user to reflect on whether all relevant parameters are included, check on whether complex code is understood (especially useful if code has been imported from other sources and may impact critical aspects of operation), or the broader work context. Critically, it offers options to ignore the suggestion, schedule it for later, or change proficiency settings (i.e., self-confidence). 

\subsection{Reducing metacognitive demands}\label{subsec:reducingmc}
In addition to improving users’ metacognition during their interaction with GenAI, systems should be designed to \textit{reduce their metacognitive demands}. Target areas for this include the explainability and customizability of GenAI systems.

\subsubsection{Explainability}\label{subsubsec:explain1}
We adhere to the definition of explainability as that which enables \textit{``people’s understanding of the AI to achieve their goals''} \cite{sun_investigating_2022}. To this point, §\ref{sec:demands} demonstrated how users struggled to understand GenAI systems and achieve their goals due to GenAI’s metacognitive demands. We focus on those using GenAI systems as tools in their workflow (in contrast to, e.g., those engaging solely with GenAI system outputs, or overseeing regulatory aspects), in line with the context-specificity of explainability advocated by research on HCXAI \cite{liao_ai_2023,suresh_beyond_2021,ehsan_human-centered_2020}. 

From the perspective of metacognitive demand, by providing contextual and performance information alongside the system inputs and outputs, explainability should help partly \textit{offload} metacognitive processing from users' minds and onto the system interface. As we illustrate below, explainability approaches can surface the information necessary for adjusting confidence in prompting, output evaluation, and automation strategy. Moreover, by providing actionable information, explainability can enable users' self-awareness and metacognitive flexibility \cite{mansi_why_2023}.\footnote{In this sense, explainability can also be viewed as a metacognitive support strategy as per §\ref{subsec:improvingmc}.} 

Explainability for GenAI systems should help users adjust their confidence in their ability to prompt, evaluate outputs, and determine their automation strategy, particularly given the multiple, non-intuitive failure modes common to GenAI systems, and other challenges \cite{liao_ai_2023}. For example, explanations that map each aspect of an output to aspects of the prompt (e.g., using attention visualization \cite{tang_what_2022}), and compare this to examples of effective prompts \cite{brachman_follow_2023,jiang_discovering_2022}, can help users disentangle issues with their prompt from those stemming from model performance, thereby supporting confidence adjustment for prompting ability. Likewise, `co-auditing' a GenAI system by revealing the model’s step-by-step actions (e.g., in spreadsheet software \cite{liu_what_2023}) can enable users to understand exactly what’s involved in a longer workflow \cite{gordon_co-audit_2023}, and help them adjust their confidence in their evaluation ability. For example, an explanation that introduces domain concepts or terminology unfamiliar to users can signal insufficient domain expertise to evaluate this output \cite{simkute_experts_2020}, and prompt further explanation. Finally, indicating model uncertainty \cite{bhatt_uncertainty_2021, sun_investigating_2022,weisz_perfection_2021}, for example, by means of color-coding \cite{spatharioti_comparing_2023}, and an explanation for that uncertainty \cite{agarwal_quality_2021,vasconcelos_generation_2023}, can also help users disentangle the role of their prompting and output evaluation ability from that of models' capabilities, further supporting confidence adjustment \cite{weisz_toward_2023}. %

The broader user workflow that encompasses the local interaction with GenAI constitutes an important usage context for explainability (see also \cite{lai_towards_2023,sun_investigating_2022}). To this end, global explanations about model capabilities for a given task can help users adjust their confidence in their ability to complete the task manually versus with GenAI support (i.e., determining their automation strategy) \cite{jiang_discovering_2022}. Indeed, for AI-assisted coding, users requested information about overall output quality and runtime performance \cite{sun_investigating_2022}.

Explainability can also reduce the metacognitive demand for user self-awareness and metacognitive flexibility. For example, explanations about effective prompting strategies for a given task—most frequently requested by users in a GenAI-assisted coding system \cite{sun_investigating_2022}—can help users translate their goals into actionable prompts or flexibly adjust their prompts (which can equally be viewed as supporting their metacognitive abilities, as in feedforward design \cite{vermeulen_crossing_2013}). Moreover, granular model uncertainty estimates, such as line-level highlighting of generated code, can support users in prioritizing their output evaluation \cite{weisz_perfection_2021,vasconcelos_generation_2023,weisz_better_2022}, thereby enabling metacognitive flexibility. In sum, these approaches to explainability all aim to reduce the metacognitive demand of GenAI systems, enabling users to take concrete actions, including `mental state' actions (e.g., confidence adjustment), system interactions (e.g., updating prompts), or actions external to the system (e.g., completing a task without GenAI) \cite{mansi_why_2023}.

As suggested in §\ref{subsubsec:selfeval}, the above explainability approaches can be augmented via metacognitive self-evaluation interventions that encourage self-awareness. GenAI offers a unique opportunity to further augment these interventions through interactivity, as advocated in recent work \cite{langer_what_2021,miller_explanation_2019,wang_designing_2019,arya_one_2019,lakkaraju_rethinking_2022}, including for GenAI specifically \cite{sun_investigating_2022}. Interactivity could be especially important for GenAI explainability, as due to GenAI's model flexibility, generality, and originality, users’ explanation needs will be as diverse as their use-cases, outputs, and metacognitive abilities.

\subsubsection{System customizability}\label{subsubsec:custom}
The question of how much control to give users---system customizability, or how many `knobs' a user can adjust---is another design choice that can moderate the metacognitive demands of GenAI \cite{dang_ganslider_2022,mcnutt_design_2023}. 

On one hand, increasing customizability can increase the demand for self-awareness (e.g., \textit{``are any of the settings relevant to my task?''}), well-adjusted confidence (e.g., \textit{``did any of the settings affect the output quality, or is it related to my prompting?''}), task decomposition (e.g., \textit{``what is the right order to adjust settings for each of my sub-tasks?''}), and metacognitive flexibility (e.g., \textit{``which settings should I adjust to improve my output, if any?''}). This may increase cognitive load, particularly for novice users. Indeed, in \citet{perry_users_2022}, half of the users did not adjust any model parameters, even though many produced insecure code using an AI assistant. To this end, the \textit{Prompt Middleware} framework aims to reduce the demand to craft prompts from scratch, enabling users to choose limited customizability \cite{macneil_prompt_2023}.

On the other hand, increased customizability can support metacognition, particularly for more advanced users. For example, consider the temperature setting, which determines the extent of non-determinism in GenAI outputs, and therefore the likelihood of hallucinations. Allowing users to change this setting can support more flexible and self-aware problem-solving in experienced users, by presenting them with different and perhaps surprising perspectives \cite{perry_users_2022}. Allowing users to find a task-appropriate temperature setting that keeps the right balance between diversity and factuality of output, or constraining factuality in different ways (i.e., through automated post-processing and deterministic fact-checking) could therefore enable metacognitive flexibility and self-awareness. Customization may therefore increase users’ trust in and satisfaction with output~\cite{openai_customizing_2023}.  

Other settings include the size of the shortlist from which output is sampled, as well as the size of the output itself. Increasing the size of the output window may demand better-adjusted confidence to evaluate and integrate more information. However, it can also support metacognition—by adjusting these parameters, users can work on understanding which part of the output should be used and how, potentially also increasing explainability. Initial self-reports such as \cite{ruskov_grimm_2023} suggest that in order to find the optimal system settings for a task, users enter an interactive feedback loop with models, in which they clearly have to formulate their goals (promoting self-awareness), adjust their confidence in output evaluation, and flexibly adapt their workflow. 

Where the balance lies between too much and not enough customizability needs exploring \cite{chen_next_2023,weisz_toward_2023}. Combining increased customizability with metacognitive support strategies (e.g., planning, self-evaluation, self-management) is a promising direction for further research.

\begin{table*}
\small
  \caption[]{Open research questions for addressing the metacognitive demands of GenAI}
  \label{tab:rqs-opportunities}
\renewcommand{\arraystretch}{1.4}  \begin{tabular}{>{\raggedright}p{76pt}>{\raggedright}p{205pt}>{\raggedright}p{196pt}}
    \toprule
    \textbf{Area} & \textbf{Research questions} & \textbf{Suggested approaches} \cr
    \midrule
    Supporting users' metacognition & How can GenAI systems increase users' self-awareness and task decomposition during prompting?  & Explore self-reflection prompts, task decomposition support, open-ended exploration, feedforward design, and other planning interventions. \cr
      & How can GenAI systems incorporate self-evaluation interventions to support users in increasing their self-awareness and adjusting their confidence? Does this affect their automation strategy? & Explore proactive probing of users' uncertainty, prompting users to self-explain and reflect interactively, and outputting systems' confidence.\cr
     & How can GenAI systems incorporate self-management interventions to support users in determining their automation strategy and improving their self-regulation and metacognitive flexibility? & Explore automated task decomposition, detection of users' states, dynamic adaptation of output complexity, and prompting towards more structured and interactive usage of GenAI. \cr
    \hline  
    Reducing metacognitive demand & How can explainability reduce the metacognitive demand of GenAI, and what is the impact of interactive explanations? & Explore impact of surfacing interactive output explanations at different levels of complexity. \cr
      & How can understanding users' metacognitive abilities when working with GenAI systems advance approaches to explainability and updating of mental models? & Monitor metacognitive abilities during GenAI interactions and explore whether metacognitive interventions improve mental model updating. \cr
      & What is the optimal balance for GenAI system customizability to reduce the metacognitive demand, and how can it be combined with metacognitive support strategies? & Explore different levels of customizability across tasks and user proficiency levels, and their impact on task performance and metacognition. \cr
    \hline
    Managing cognitive load while addressing metacognitive demands & How do metacognitive interventions affect cognitive load as users learn to interact with GenAI systems over time, and how should interventions optimally adapt or fade out? & Explore reducing or otherwise adapting interventions at different timescales as metacognitive proficiency and task performance increases.\cr
      & What are other ways to optimize the balance between addressing metacognitive demands and overall cognitive load? & Explore context-appropriate gamification of metacognitive interventions.\cr
    \bottomrule
  \end{tabular}\renewcommand{\arraystretch}{1}
\end{table*}

\subsection{Managing cognitive load while addressing metacognitive demands}\label{subsec:cogload}
There is a risk that strategies to address metacognitive demands may also increase cognitive load, due to the additional information that users have to process, such as self-reflective prompts, a set of sub-goals resulting from task decomposition, or model explanations \cite{de_bruin_synthesizing_2020}. The relationship between metacognition and cognitive load is an active research area \cite{de_bruin_synthesizing_2020,seufert_interplay_2018,wang_cognitive_2023,wang_examining_2023}, but it is plausible that, although cognitive load may increase due to the \textit{processing requirements} of many metacognitive support interventions, the improvement in metacognition may be accompanied by a simultaneous and larger reduction in cognitive load, resulting in a \textit{net decrease} in cognitive load. Some studies show that metacognitive support does not increase overall cognitive load \cite{zheng_effects_2019,lopez-vargas_students_2017} (see also \cite{wang_cognitive_2023}). Most importantly, we hypothesize that improved metacognition should result in a net improvement in output quality.  

Likewise, we propose that explainability, by partly offloading metacognitive processing from users 
 and onto the system, should reduce the cognitive load \textit{associated with metacognitive monitoring and control}. However, it may increase the cognitive load \textit{associated with processing explanations} \cite{de_bruin_synthesizing_2020}. To the latter point, some interactive explanations have been found to increase cognitive load \cite{bertrand_selective_2023}. Ultimately, however, as users adapt to working with explainable GenAI systems, we hypothesize that the result should be a net reduction in cognitive load \cite{nuckles_self-regulation-view_2020}. As noted above, system customizability involves a similar tension between cognitive load and metacognition.

Training effects over time may be key, as users may gradually internalize the metacognitive strategies, explanations, or customization settings, and no longer need to rely on external prompts \cite{de_bruin_synthesizing_2020}. Accordingly, metacognition-related cognitive load should decrease, although it is less clear whether cognitive load associated with the \textit{processing} of external prompts also sufficiently decreases. To this point, \cite{nuckles_expertise_2010} found that adaptive and gradual fading out of metacognitive prompts produced the largest performance benefits, as it provided time for students to internalize metacognitive strategies, while ultimately reducing the cognitive load associated with processing now-irrelevant external prompts (see also \cite{nuckles_self-regulation-view_2020}). The same might be true for some types of explanations as well (e.g., global explanations). Future research should study how metacognitive interventions affect cognitive load as users learn to interact with GenAI systems over time, and how to optimally adapt or fade out interventions over time. It is also important to explore other ways to optimize the balance between addressing metacognitive demand and overall cognitive load (e.g., through gamification \cite{simkute_experts_2020}).

Lastly, and perhaps somewhat controversially, we highlight the value of `seamfulness' in interface design for helping users reflect on their technology use (as per §\ref{subsubsec:selfeval} and \cite{inman_beautiful_2019, weiser_creating_1994}). This idea can be extended to question the `doctrine of simplicity', which assumes that interfaces should always be `easy' or `natural' to use \cite{sarkar_should_2023}. We propose that \textit{some} potential effort introduced by metacognitive support strategies and explanations may be justified, so long as these are well-designed and act ultimately in the service of improved metacognition and productivity with GenAI, a technology which promises to transform personal and professional work \cite{sarkar_should_2023}.

\section{Conclusion}\label{sec:conclusion}
 Russell \cite{design_lab_what_2017} proposed that being literate in the ``Age of Google'' required a kind of ‘meta-literacy’---knowing how to read the search interface, how to use it effectively, and what is even possible to search for. Analogously, as we offload more of our cognition to today's GenAI systems, the demand for our metacognition increases \cite{risko_thinking_2023}. Designing truly human-centered GenAI systems \cite{chen_next_2023,liao_ai_2023} means grappling with these metacognitive demands. Fortunately, a rich body of metacognition and cutting-edge HCI research can kickstart this effort. Equally, interaction with GenAI offers a powerful paradigm for advancing our foundational understanding of metacognition, paving the way for fruitful inter-disciplinary research. Finally, we reiterate that the perspective of metacognition, when considered with the unique features of GenAI---model flexibility, generality, and originality---presents an opportunity to realize what Alan Kay proposed as a \textit{``grand collaboration''} with \textit{``agents: computer processes that act as guide, as coach, and as amanuensis''} 
\cite{kay_user_1990}.

\begin{acks}
We thank Siân Lindley, Andy Gordon, Sam Gilbert, and the anonymous reviewers for their constructive comments.
\end{acks}

\bibliographystyle{ACM-Reference-Format}
\bibliography{CHI2024_metacognition}

\end{document}